\documentclass[10pt,sigconf,letterpaper]{acmart}
\renewcommand\footnotetextcopyrightpermission[1]{} 
\usepackage{epsfig}
\usepackage{endnotes}
\usepackage{a4wide}
\usepackage{amsmath}
\usepackage{amssymb}
\usepackage{amsfonts}
\usepackage{amsthm}
\usepackage{enumerate}
\usepackage{graphicx}
\usepackage{caption}
\usepackage{subcaption}
\usepackage{comment}
\usepackage{verbatim}
\usepackage{multirow}
\usepackage{adjustbox}
\usepackage{changepage}
\usepackage{color}
\usepackage{time}

\usepackage{listings}

\usepackage[ruled]{algorithm}

\usepackage[noend]{algpseudocode}
\algnewcommand\algorithmicforeach{\textbf{for each}}
\algdef{S}[FOR]{ForEach}[1]{\algorithmicforeach\ #1\ \algorithmicdo}
\algrenewcommand\algorithmicindent{0.5em}%

\AtBeginDocument{%
  \providecommand\BibTeX{{%
    \normalfont B\kern-0.5em{\scshape i\kern-0.25em b}\kern-0.8em\TeX}}}
\setcopyright{none}
\begin{document}
\newcommand{\mo}{\mathbb{O}}
\newcommand{\mf}{\mathbb{F}}
\newcommand{\mb}{\mathbb{B}}
\newcommand{\mv}{\mathbb{V}}
\newcommand{\MO}{$\mathbb{O}$}
\newcommand{\MOs}{$\mathbb{O}$s}
\newcommand{\MMO}{malicious $\mathbb{O}$}
\newcommand{\MF}{$\mathbb{F}$}
\newcommand{\MFs}{$\mathbb{F}$s}
\newcommand{\MMF}{malicious $\mathbb{F}$}
\newcommand{\MB}{$\mathbb{B}$}
\newcommand{\MBs}{$\mathbb{B}$s}
\newcommand{\MMB}{malicious $\mathbb{B}$}
\newcommand{\MV}{$\mathbb{V}$}
\newcommand{\DE}{digital exchange}
\newcommand{\FDE}{\emph{fair exchange}}
\newcommand{\MBC}{$\mathcal{L}$}
\newcommand{\MSC}{$\mathcal{C}$}
\newcommand{\MCHAN}{$\mathcal{\zeta}$}
\newcommand{\MCHANID}{\MCHAN.$cid$}
\newcommand{\MCHANST}{\MCHAN.$Store$}
\newcommand{\MCHANLD}{\MCHAN.$Load$}
\newcommand{\MCHANSEND}{\MCHAN.$Send$}
\newcommand{\MCHANRECV}{\MCHAN.$Recv$}
\newcommand{\MCHANCLOSE}{\MCHAN.$Close()$}
\newcommand{\vader}{VADER}
\newcommand{\BME}{{\it Blockchain Mediated Exchange}}
\newcommand{\fullon}{BME}
\newcommand{\vanilla}{VANILLA}

\newcommand\blfootnote[1]{%
  \begingroup
  \renewcommand\thefootnote{}\footnote{#1}%
  \addtocounter{footnote}{-1}%
  \endgroup
}

\title[VADER]{Verifiable and Auditable Digital Interchange Framework\textsuperscript{\#}}
\author{Prabal Banerjee$^\dagger$,
        Dushyant Behl$^\ast$,
        Palanivel Kodeswaran$^\ast$,
        Chaitanya Kumar$^\ast$,
        Sushmita Ruj$^\ddagger$ and
        Sayandeep Sen$^\ast$}
\affiliation{%
  \institution{$^\dagger$Indian Statistical Institute Kolkata, $^\ast$IBM Research, $^\ddagger$ CSIRO Data61 Australia and Indian Statistical Institute Kolkata}
}

\begin{abstract}
We address the problem of fairness and transparency in online marketplaces selling digital content, where all parties are not actively participating in the trade. We present the design, implementation and evaluation of \vader, a highly scalable solution for multi-party fair digital exchange that combines the trusted execution of blockchains with intelligent protocol design and incentivization schemes. We prototype \vader\ on Hyperledger Fabric and extensively evaluate our system on a realistic testbed spanning five public cloud datacenters, spread across four continents. Our results demonstrate that \vader\ adds only minimal overhead of 16\% in median case compared to a baseline solution, while significantly outperforming a naive blockchain based solution that adds an overhead of 764\%.
\end{abstract}

\maketitle
\blfootnote{\# Authors are listed in alphabetical order.}
\section{Introduction}
Online media consumption is a big business~\cite{prsnews}, with users watching billions of hours of videos per
month~\cite{YoutubeStats} and media traffic constituting roughly 70\% of downstream internet traffic
~\cite{sandvine2018}.

\begin{figure}[t]
 \centering
	\includegraphics[width=0.9\columnwidth]{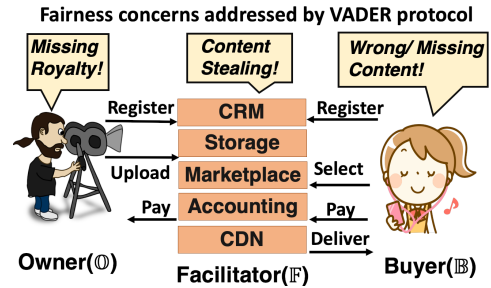}
	\vspace{-0.2cm}
 \caption{Present day video exchange}
	\vspace{-0.2cm}
 \label{fig:state_of_the_art}
 \end{figure}

A key reason for this success, lies in the simplicity of present day online media (and money) exchange process as depicted in Fig.~\ref{fig:state_of_the_art}. As shown, a present day content creator can simply upload content and get paid based on viewership (or sales). Similarly, buyers can pay the right price to access media without worrying about content authenticity, price gouging, non delivery etc.
The ease of operation is due to presence of facilitators such as Youtube, Netlflix, iCloud etc. As shown in Fig.~\ref{fig:state_of_the_art}, the facilitator provides all the ancillary but critical services of content hosting, searching, delivery, payments etc. to complete the digital ecosystem for online media consumption.
While the efficacy of present day media delivery systems is reflected in their success, they suffer from an important shortcoming in the {\it inability to guarantee honest behavior} and vulnerability to fraudulent behavior by participants. This is due to the fact that present day systems lack the underpinnings to demonstrably guarantee honest behavior; forcing both buyers and owners to {\it trust} that a facilitator behaves in a {\it fair} manner.
\label{sec:intro}

For example in Fig.~\ref{fig:state_of_the_art}, the owner trusts the facilitator to honestly report the number of views/downloads and calculate its royalty payments in a transparent manner. Similarly, the buyer trusts the facilitator to deliver the right content against payment. Finally, the facilitator expects the buyer to pay for a successful delivery, without falsely alleging non-delivery. 

However, lack of baked-in guarantees of honest behavior can lead to disputes, such as a buyer fraudulently alleging non-receipt of content and denying payments thus stealing content from facilitator; facilitator mis-reporting sales to cheat owner of dues, or even charging buyers without providing right content. 

In fact, recent events highlight the inadequacy of this trust based model~\cite{YoutubeTransparency,YoutubeVlogger,YoutubeMonetizationPolicyChanges}. Specifically, content owners have raised a number of complaints against faclitators regarding their royalty earnings and lack of clarity in the calculation~\cite{YoutubeFraud,spotify_royalty_controversy,YoutubeRevenueDrop}, highlighting discrepancies in earnings with reported viewership. Similarly, buyers have also raised disputes regarding content received from these platforms~\cite{AppleReturnFunda}.


We believe that such disputes can only increase in future to the
detriment of the growth of online media delivery platforms. This in turn, motivates us to address this important problem of {\it guaranteeing} successful video exchange amongst mutually untrusting buyers and facilitators. We ensure each party (owner, buyer and facilitator) gets their rightful outcome or no one does. We define the above as \textbf{Multi-party fair digital exchange} (abbreviated as \FDE\
in the rest of the paper) among the owners, buyers and facilitators in the marketplace model. 

Note that prior work~\cite{adtributor,auditable_vms,peerreview} for providing fairness are not applicable in the above setting as they {\it trust the facilitator} to grant access to internal logs and metrics, an important assumption we intend to avoid. 
Similarly, decentralized video delivery platforms~\cite{LBRY,livepeer,vivuly,bittube,dtube} which bypass centralized facilitators (thus obviating the need to {\it trust} them), while promising are not suitable.
This is due to the fact that decentralized content delivery platforms suffer from poor content quality, sporadic availability, rampant illegal content etc., ironically due to absence of a facilitator's dedicated resources in maintaining the platform~\cite{UnreelingNetflix}.

Motivated by above observations, we set our goal to develop a {\it readily usable} solution for \FDE, which would be a){\it compatible and incrementally deployable with present day facilitator driven marketplace systems} and b) {\it should be able to provide transparency and fairness into existing video delivery platforms with minimal overhead in terms of modifications and performance}.

In this paper we present the system design, implementation and evaluation of Verifiable and Auditable Digital Interchange Framework (VADER) which satisfies the above mentioned criteria. 

In process of designing \vader, we study various fraud risks that arise in the marketplace model and note that guaranteeing multi-party \FDE\ in this model presents a unique challenge. As shown in Fig.~\ref{fig:state_of_the_art}, the owner is a {\it passive party}, in the sense that after video upload, it is not directly involved in the exchange of video and money between the {\it active parties} viz. the facilitator and buyer. Being a passive party, an owner is completely dependent on the honesty of facilitator, as it has no way to learn of exchanges of its content being done. This makes the owner vulnerable to being misled about its true earnings, either by the facilitator~\cite{YoutubeFraud} alone, or in collusion with the buyer~\cite{AppleReturnFunda,YoutubeCopyrightScam}. 
Fig~\ref{fig:state_of_the_art}, highlights the specific risks faced by the respective parties. 

We note that while the problem of fair exchange among active parties is well studied in theory~\cite{Rabin1981HowTransfer,Goldreich:1987:PAM:28395.28420,YaoExchangeSecrets,FairExchangeInfeasibility}; to the best of our knowledge, protecting rights of passive parties, without significantly altering the flow of video exchange~\footnote{i.e., without making owner also an active party by say asking for its approval on every trade} {\it is a new paradigm for fair exchange, not yet covered in literature.}

In \vader\ we not only protect the buyer and facilitators against various {\it active party risks} but secure owner's interest from {\it passive party risks}. To the best of our knowledge, \vader\ is the first real system to demonstrate such capabilities.

\vader\ accomplishes low overhead \FDE\ solution by leveraging the following key insights,
\newline
$\rightarrow$\textbf{Insight 1)} We can {\it guarantee} \FDE\ by sending encrypted video and performing
fair exchange of only the {\it key} and money. This
enables us to leverage the existing optimized delivery infrastructures
of facilitators for sending (encrypted) content and making system for fairness incrementally deployable.
\newline
\noindent$\rightarrow$\textbf{Insight 2)} By assuming the presence of a trusted arbitrator that is slow (when compared to direct interaction without intermediary)
but can deterministically detect a malicious party and provide restitution (right encrypted content, key or money) to
the honest party, parties can opportunistically exchange key and
money directly between themselves without having to interact with the
arbitrator unless there is a dispute. 
\newline 
\noindent$\rightarrow$\textbf{Insight 3)} Assuming parties are rational, introduction of bounties (that are
large and funded by penalizing malicious parties) for reporting misbehavior introduces an element of distrust between
parties, thus preventing collusion aimed at subverting the protocol. 
Note that, the first two insights enable efficient operation, guaranteeing fairness for the active parties; while the third insight ensures fairness for the passive party under assumption of rational participants.

We select blockchain as the tamper-proof ledger and execution platform for \vader. Our selection of blockchain is motivated by the fact that it offers decentralization of trust and auditability guarantees sought by \vader. Furthermore, the native blockchain cyptocurrency can be used to design incentivization schemes and programmatically enforce desired behavior from the interacting parties as mandated by our insights above.
We also point out the second insight of opportunistic exchanges on fast path (batching), while reverting to the slow but guaranteed path is also used in state-channels for scaling transaction throughput. However, these solutions involve native assets such as cryptocurrencies giving complete control to the arbitrator to revert back the state (e.g. ownership) of an asset. On the other hand, we deal with non-native assets such as decryption keys which once delivered to the buyer are unaffected by blockchain asset state. We modify the state-channel protocol to account for the above oddity, as described in Sec.~\ref{sec:outline}.

As part of this work, we have systematically studied exchange process in present day video delivery platforms and used the insights to design and implement \vader. \vader\ protocol carefully combines insights from diverse domains of cryptography, incentives design, blockchains to ensure \FDE\ for video exchange. \vader\ is incrementally deployble with minimal modifications to present day video delivery platforms. 
Specifically, in Sec.~\ref{sec:solution} we design \vader\ protocol (message exchange
sequence) 
and perform a comprehensive security analysis to
show how \vader\ protects honest parties against attacks in Sec.~\ref{sec:security}. We implement
\vader\ incorporating all the insights described above and
extensively study the performance of \vader\ over two baseline
techniques, through extensive evaluation across realistic workloads.

We note that while we use video as an example digital asset in this paper, \vader\ works for any digital content~\cite{spotify,etsy,GooglePlayStore,steam} that can be provably verified, say using cryptographic digests. Additionally, the owner in \vader\ is a logical abstraction that can represent multiple entities that need to be paid royalties individually, as in the music industry~\cite{spotify_royalty_controversy}. Finally, \vader\ focuses on guaranteeing \FDE\ between the buyer, facilitator and owner. \vader\ does not address the complementary issue of content piracy, where a buyer buys content legally on a \vader\ based marketplace and then resells it with minor modifications. Other works~\cite{eme_drm,widevine}, including recent work~\cite{kate_watermarking} use a combination of content watermarking and on-blockchain penalty mechanisms to prevent content piracy which can be easily embedded into \vader\ smart contracts.

We believe that \FDE\ systems will be the norm in near future and our work advances the nascent area of designing practical systems for Multi-Party Fair exchange.
We claim the following as key {\bf contributions:}
\textbf{1)} To the best of our knowledge, we are the first to
formulate and present the problem of \textit{multi-party fair digital
 exchange} in third party marketplace scenario where one of the parties is passive and does not directly interact with the buyer (Sec.~\ref{sec:intro} \& \ref{sec:setting}).
 \textbf{2)} We design the \vader\ protocol and study its security properties.
\textbf{3)} We implement \vader\ protocol on Hyperledger Fabric, and extensively evaluate its performance on a realistic test-bed of
upto 91 nodes spread over 4 continents, transferring at least 50TB of
data over the network. We find that \vader\ adds only minimal overhead of
16\% in median case compared to the baseline \vanilla\ solution.

{\bf Outline:} The rest of the paper is organized as follows. In
Sec.~\ref{sec:setting}, we formally describe the problem and the
solution requirements. In Sec.~\ref{sec:solution}, we describe our
solution and discuss the security analysis of our work in Sec.~\ref{sec:security}. We present details of \vader\ implementation in
Sec.~\ref{sec:implementation} and in Sec.\ref{sec:evaluation},
we present performance evaluation of \vader\
 under realistic
conditions. In Sec.~\ref{sec:related}, we discuss the related work
and finally conclude our paper in Sec.~\ref{sec:conclusion}

\section{Problem Setup}
\label{sec:setting}
We define Owner (\MO) as a party, such as artists or financiers, that need to be monetarily compensated for every successful trade/download of the digital content.  We define Facilitator(\MF) as a party responsible for conducting the trade on behalf of \MO. Finally, we define a Buyer(\MB) as a party that is buying digital content from \MF's\ marketplace.

As explained in Sec.~\ref{sec:intro}, \MO\ is the {\it passive party}
while \MB\ and \MF\ are {\it active parties} in the digital
exchange. Further, we assume the parties \textit{do not completely trust each other}
and \MB\ and \MO\ do not know each other. We also assume that the parties
are rational and will collude with each other to maximize profit.
\newline
\indent A \textbf{Multi-party Fair Digital Exchange} protocol in the above
context, should ensure that either {\it all} parties receive their
desired item in exchange for their own item/service, or {\it none} of
the parties receive anything. Specifically, \textbf{Multi-party Fair Digital Exchange}
guarantees the following behavior:
\textbf{a)} A \MB\ will always get the correct video uploaded by 
\MO\ if it can submit proof of payment
\textbf{b)} A \MF\ will always get paid if it can submit evidence of
sending the correct video
\textbf{c)} An \MO\ will always get paid if \MF\ was paid by selling its content.

\noindent \textbf{Attacks:} A \FDE\ protocol for video delivery must protect against the following attacks:
{\bf Atk.1- Royalty Manipulation:} constitutes any manipulation in
the royalty calculation of \MO\ e.g. a malicious \MF\ can cheat \MO\ by
a) misreporting the number of downloads or b) selling the content to
\MB\ at a higher price while compensating \MO\ with a lower royalty by
exploiting the  information asymmetry between \MB\ and \MO.
{\bf Atk.2- Content Mismatch}: constitutes delivery of content 
to a \MB\ that is different from the one promised while receiving money.
{\bf Atk.3- Content Stealing}: constitutes gaining access to the
content without compensating the designated  parties for their
fair share of payment for the same, e.g. \MB\ getting access to
content without paying money to \MF\ by claiming to have received
wrong content despite getting right content, thus cheating both \MF\ and \MO.

Under our threat model, the above attacks represent a completely exhaustive list of
attacks on marketplace fairness. We reiterate that \vader\ does not
address the complementary issue of content piracy, and is compatible
with other solutions~\cite{eme_drm,widevine,kate_watermarking} that address piracy.

\begin{figure*}[htb]
    \begin{minipage}[t]{0.6\textwidth}
	\centering
	\includegraphics[width=\columnwidth]{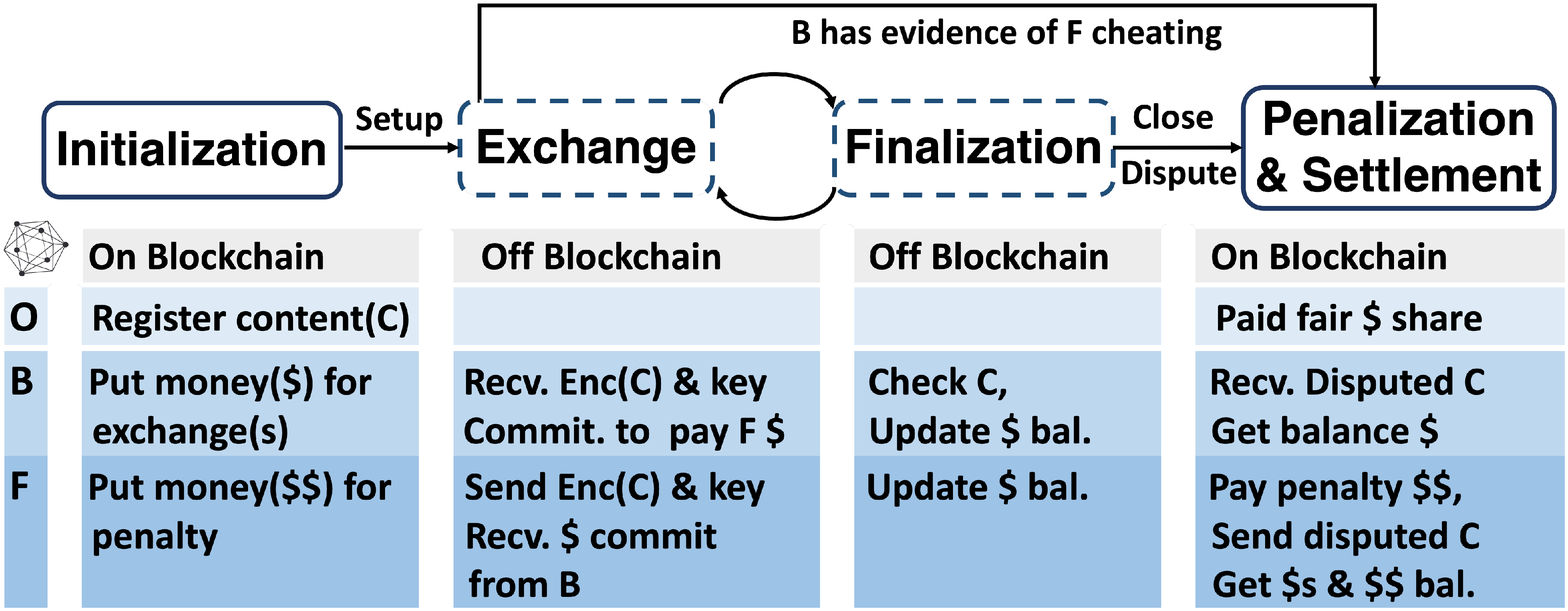}
	\caption{Video exchange lifecycle in \vader}
	\label{fig:vader_lifecycle}
    \end{minipage}
    ~
    \begin{minipage}[t]{0.4\textwidth}
 	\centering
 	\includegraphics[width=\columnwidth]{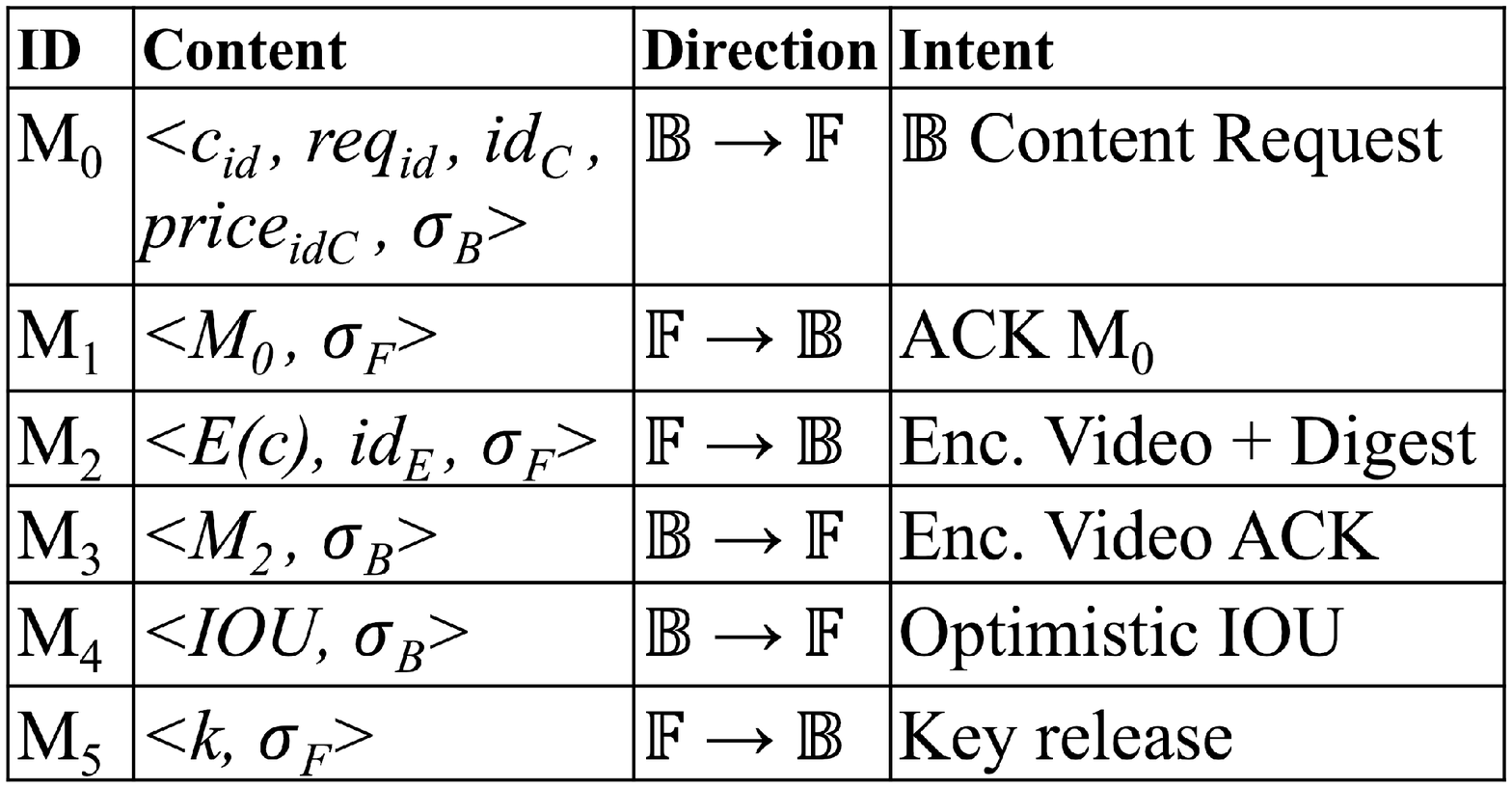}
	 \caption{\vader\ message exchanges}
 	\label{fig:sc_fsm}
    \end{minipage}
\end{figure*}

\section{Solution Description}
\label{sec:solution}
We describe \vader\ protocol by presenting an outline of the solution in Sec.~\ref{sec:outline} and the details of various protocol steps in Sec.~\ref{sec:solution_desc}. We also, present a security analysis of our protocol in Sec.~\ref{sec:security}.
\newline
\newline
\noindent \textbf{$\bullet$ Notation: } We use the following notation in the description of our solution. Let $\lambda$ be a security parameter, $H:$ $\{0,1\}^*$ $\rightarrow$ $\{0,1\}^\lambda$ be a cryptographically secure hash function, ($Gen(\lambda)$, $Enc(k,m)$, $Dec(k,m)$) be a IND-CCA secure symmetric encryption scheme and ($KeyGen(\lambda)$, $Sign(sk,m)$, $Verify(pk,m,\sigma)$) be a public key cryptosystem based secure digital signature scheme. We assume that at bootstrap, each party generates a key pair $(pk,sk)$ using $KeyGen$. We use \MBC\ to represent the distributed blockchain ledger and \MCHAN\ to represent a state channel between two parties.

\begin{algorithm}
	\caption{Conditional Escrow Contract}
	\label{alg:escrow}
	\begin{flushleft}$IOU:=<sender,recvr,\$amt>$\end{flushleft}
	\begin{algorithmic}[1]
		\Function{OpenEscrow} {$B,amt,\sigma_{B}, \tau,cond$}
	  \If{$Verify(pk_B, amt, \sigma_{B}) = true$}
		\State $e_{id} \gets genEscrow()$
		\State $e_{id}.[owner, bal, timeout, cond] \gets$
		  \State $\ \ \ \ \ \ [B, amt, \tau, cond]$
		\State {$ \Return$ $ e_{id}$}
	  \EndIf
	  \EndFunction
		\Function{$ProcessIOU$}{$[IOU(s)], evidence$}
		\State \textbf{if} {$block_{ht} \le e_{id}.timeout$ AND}
		\State \ \ $Verify(evidence,e_{id}.condition$) \textbf{then}
		  \ForEach{$iou in IOU(s)$}
			\State {$bal$ -= $iou.amt$}
			  \State {$send(iou.recvr, iou.amt)$}
		  \EndFor
	  \EndFunction
	  \Function{$CloseEscrow$}{$e_{id}$}
		\If {$block_{ht} \ge timeout$}
			\State {$send(e_{id}.owner,e_{id}.balance)$}
		\EndIf
	  \EndFunction
	\end{algorithmic}  
  \end{algorithm}
  
  \begin{algorithm}
	\caption{State Channel Contract}
	\label{apx:alg:sc}
	 \begin{flushleft}\ \ $\tau$\ represents the settlement timeout\end{flushleft}
	 \begin{algorithmic}[1]
	   \Function{Open}{\MB,$B_{amt}$,$\sigma_{B}$,\MF,$F_{amt}$,$\sigma_{F}$,$\tau$}
	   \State \MCHAN $= GenChannel()$
	   \State \MCHANID$\ \gets \{0,1\}^k$
	   \State \MCHAN$.timer\_started = false $
	   \State {\MCHAN.\MB.$escrow = OpenEscrow($\MB\,$B_{amt}$,$\sigma_{B},\tau)$}
	   \State {\MCHAN.\MF.$escrow = OpenEscrow($\MF\,$F_{amt}$,$\sigma_{F},\tau)$}
	   \State \MCHAN$.timeout$, \MCHAN$.state$ = $\tau$, $Open$
	   \State \MBC.Store(\MCHAN)
	   \EndFunction
	   \Function {Close}{\MCHAN,${ChanUpdateMsgs}$} 
	   \State \textbf{if} {\MCHAN.$timer\_started$ AND $block_{ht}$ $\ge$ $\text{\MCHAN}.timeout$}
	   \State \ \ {$ \Return  $}
	   \ForEach {$m \in ChannelUpdateMessagese $}
	   \State {$VerifyValidity(m)$}
	   \State {$iou = m.extractIOU()$}
	   \State $ProcessIOU(\text{\MCHAN}.escrow(s),iou)$
	   \EndFor
	   \State \MBC.Store({\MCHAN.$timer\_started = true$})
	   \EndFunction
	   \Function {settleChannel}{$c_{id}$}
	   \State \textbf{if} {\MCHAN.$timer\_started$ AND $block_{ht} \ge \text{\MCHAN}.timeout$}
	   \State \ \ {$CloseEscrow(\text{\MCHAN}.escrow(s))$}
	   \State \MBC.Store({\MCHAN$.state = Closed$})
	   \EndFunction
	\end{algorithmic}
  \end{algorithm}

\subsection{Solution Outline}
\label{sec:outline}
\vader\ aims to deliver an efficient solution for guaranteeing \FDE\ by leveraging the three insights described in Sec.~\ref{sec:intro}. 
Following the first insight, \MF\ utilizes its usual content delivery infrastructure to send encrypted
video to \MB\, thereby reducing the problem of fair video exchange to that of fair exchange of key and money.

Fig.~\ref{fig:vader_lifecycle} depicts the phases of video
exchange using \vader. In the Initialization phase, leveraging the second insight,
\MF\ and \MB\ lock
money in the blockchain escrow to create a channel, akin to
state-channels~\cite{poon_lightning, statechannelnetworks}. We define
\textit{channel} as a unique bi-directional message queue between two
parties, backed with blockchain~escrowed money, enabling parties to
conduct multiple exchanges directly without hitting blockchain
(Channel construction and lifecycle described later). This money is then used as collateral
in the Exchange and Finalization phases where the \MB\ and \MF\
exchange money and video directly offchain. Finally, in the
Penalization and Settlement phase, disputes, if any, are resolved
deterministically by the trusted arbitrator on-chain as shown in
Fig.~\ref{fig:vader_lifecycle} and escrow money appropriately
redistributed to \MB\ and \MF. We note that, as shown in Fig.~\ref{fig:vader_lifecycle} only Initialization and Settlement phases interact with blockchain, while exchange and finalization phases are executed directly between \MB\ and \MF\ without hitting the blockchain. Moreover, \MO\ being the passive party does not have any actions during the exchange and finalization steps of the protocol.

\vader\ leverages the trusted execution of a suite of blockchain deployed smart contracts, to emulate a {\it trusted arbitrator}.
The trusted arbitrator (in settle phase) works as follows: 
\textbf{1)} Any party can close the channel either due to dispute or
logical end of application exchange, and submit evidence of protocol
compliant behavior to the smart contract, if any
\textbf{2)} \vader\ allows the counterparty to submit counterfactual
evidence within a pre-determined time period to state its view of the
offchain protocol state.
As shown in Fig.~\ref{fig:vader_lifecycle}, 
the \vader\ smart contracts evaluate the evidence submitted by both parties and then either transfer money to \MB\ (refund) or \MF\ (payment) from an escrow that is funded in the initialization phase. 

Finally, motivated by the third design insight, \vader\ protocol design introduces a bounty scheme that rewards \MB\ with a much larger monetary benefit, by penalizing \MF\ (compared to cost of the video) if it can submit evidence of cheating by \MF\ to the arbitrator. 
As mentioned before, under assumptions of rationality, the penalty
scheme prevents collusion between \MF\ and \MB\ in which they both agree to
alter the offchain message exchanges between themselves for mutual
benefit (by dividing \MO's share between themselves), and submit the altered
sequence in the settlement phase, thereby cheating \MO\ of its fair share.
By making it more beneficial for \MB\ to report \MF\ and collect bounty, in turn, forcing \MF\ to honestly report exchange information to
the arbitrator, \vader\ ensures the passive party, \MO, gets its rightful share.    
Finally, we note that \vader\ uses two well known constructs described below.

\noindent\textbf{Evidence of protocol compliance}: 
\vader\ leverages cryptographic commitment schemes and digital
signatures over  messages exchanged between \MF\ and \MB\ as {\it evidence} of protocol compliant behavior by blockchain based trusted arbitrator.

\noindent\textbf{Conditional Escrow}:
This construct allows parties to prove solvency to each other as the first step of digital exchange by locking money on blockchain and guaranteeing that the money cannot be unilaterally withdrawn by either party. This construct is used by the {\it trusted arbitrator} to lock money (crypto-currency) for a duration ($\tau$) in the initialization phase and deterministically release money ($IOU$) based on submission of evidence ($cond$) for \vader\ protocol compliant operation in settle phase. Alg.~\ref{alg:escrow}, shows a conditional escrow contract, wherein an entity can lock money($amt$), for a duration of time($\tau$). Money($IOU$) is releasable before $\tau$ only on successful evaluation of condition $cond$.
Alg.~\ref{alg:escrow} in Appendix shows pseudocode for conditional escrow contract.
\newline
Our state channel construction builds on top of the conditional escrow contract. In Alg.~\ref{apx:alg:sc} we provide our state channel algorithm providing open, close and settle semantics. In addition to the one's in Alg.~\ref{apx:alg:sc} our state channel supports the following properties (used in subsequent algorithms), \textbf{1)} \MCHAN.Store and \MCHAN.Load:- which is the per party local channel storage where each party can save messages,
\textbf{2)} \MCHAN.Send and \MCHAN.Recv:- using which one party can send a message to the other party involved in the state channel.
\subsection{Solution Description}
\label{sec:solution_desc}
As mentioned in Sec.~\ref{sec:outline}, \vader\ proceeds in the four phases as depicted in Fig.~\ref{fig:vader_lifecycle}. We describe the specifics of each of the phases next.

\begin{algorithm}[t]
  \caption{Content Registration}
  \label{apx:alg:upload}
  \begin{flushleft}\ \ $amt_O\ \triangleright$ \% Royalty \MO\ should receive per Xchg \\\end{flushleft}
  \begin{algorithmic}[1]
    \Function{\MO.UpContent}{$C,n,amt_O$}
      \State $c_1, c_2, \dots, c_n = C$ \;
      \State $id_C = \{H(c_1),H(c_2),\dots,H(c_n)\}$ \;
      \State $m = <id_C, amt_O, pk_O, pk_F>$ \;
      \State Send <$C$, $m$, $\sigma_O$ = $Sign$($sk_O$,$m$)> to \MF\
      \State Receive $<Ack, vid>$ from \MF\;
    \EndFunction
  \end{algorithmic}
  \begin{algorithmic}[1]
    \Function{\MF.RecvContent} {\MBC}
      \State On Recv $<C, m, \sigma_O>$ from \MO\;
      \If{$Verify(pk_O, m, \sigma_O) \neq true$}\;
        \State terminate \EndIf 
      \State $id'_C = \{H(c_1),H(c_2),\dots,H(c_n)\}$\;
      \State $<id_C, amt_O, pk_O, pk_F> = m$\;
      \If{($m.id_C \neq id'_C)$}\;
        \State terminate \EndIf
      \State $vid = H(H(c_1)|H(c_2)|\dots|H(c_n))$\;
      \State $\sigma_F =  Sign(sk_F, <m, \sigma_O>)$\;
      \State \MBC.$Store(vid, m, \sigma_O, \sigma_F)$\;
      \State Send $<Ack, vid>$ to \MO\;
    \EndFunction
  \end{algorithmic}
\end{algorithm}

\begin{algorithm}[t]
  \caption{Trade Agreement}
  \label{apx:alg:trade_req}
  \begin{flushleft}\ \ $id_C\ \triangleright$ The content which \MB\ is requesting \\
  \ \ $price_{id_C}\ \triangleright$ Bid price \MB\ wants to pay \\
  \ \ $cost_{id_C}\ \triangleright$ Ask price which \MF\ wants \\
  \end{flushleft}
  \begin{algorithmic}[1]
    \Function{\MB.ContentReq}{\MCHAN, $id_C, price_{id_C}$}
      \State $reqid \leftarrow \{0,1\}^k$;
      \State $m = <\text{\MCHANID}, reqid, id_C, price_{id_C}>$\;
      \State $\sigma_B = Sign(sk_B, m)$;
      \State \MCHANSEND$(M_0 <m,\sigma_B>)$ to \MF;
      \State $\triangleright$ $M_1$ is generated in ServContentReq
      \State $M_1 <m,\sigma_B,\sigma_F>$ = \MCHANRECV() from \MF\;
      \State \MCHANST($M_0, M_1$)\;
    \EndFunction
  \end{algorithmic}
  \begin{algorithmic}[1]
    \Function{\MF.ServContentReq} {\MCHAN, \MF, \MB}
      \State $M_0 <m, \sigma_B>$ = \MCHANRECV() from \MB\;
      \If{$Verify(pk_B, m, \sigma_B) \neq true$}\;
        \State \MCHANCLOSE \EndIf
      \State $<reqid, id_C, price_{id_C}> = m$\;
      \If{$(price_{id_C} \neq cost_{id_C})$}
        \State \MCHANCLOSE \EndIf
      \If{($reqid$ is prev. known in \MCHAN)}
        \State \MCHANCLOSE \EndIf
      \State $\sigma_F = Sign(sk_F, (m, \sigma_B))$\;
      \State $M_1 = <m,\sigma_B,\sigma_F>$\;
      \State \MCHANSEND($M_1$) to \MB\;
      \State \MCHANST($M_0, M_1$)\;
    \EndFunction
  \end{algorithmic}
\end{algorithm}

\noindent\textbf{$\bullet$ Phase 1: Initialization:}
This phase involves \MO\ uploading content to the \MF\ and \MB\
setting up a channel to be used for multiple exchanges with \MF.
\newline
\noindent\textbf{{Init.1} \MO\ - \MF\ Content Registration:}
As shown in Alg~\ref{apx:alg:upload}, Owners register their content for sale in the marketplace by uploading
the content $C$(=$\{c_{1..n}\}$) composed of chunks $c_1,...c_n$
along with its digest $id_C$=$\{H(c_1)$,$\dots$,$H(c_n)\}$ and $amt_O$, the
percentage to be given to \MO\ for each sale of $C$ (for privacy reasons $amt_O$ is
encrypted with the public key of \MF). 
On receiving $C$, \MF\ verifies the message signature 
and records \MO\ as the unique owner and \MF\ as an authorized distributor on blockchain. Note that content registration is a
{\it one time} operation between the \MO\ and \MF. At the end of this step, the rightful ownership of video
content is established on blockchain.

\noindent \textbf{{Init.2} \MB\ - \MF\ Channel Creation:}
During initialization, \MB\
funds the channel (with \MF) with crypto-currency (sufficient for multiple
video exchanges) by locking money in an on-chain escrow. Similarly,
\MF\ also makes an initial deposit, that is much larger compared to \MB\'s
deposit (to cover penalties described later). 
Once both deposits are committed to blockchain, the state
channel contract generates a unique channel id \textit{$cid$} that is used to bind
subsequent message exchanges between \MB\ and \MF\ to the channel.
\newline
\noindent\textbf{$\bullet$ Phase 2: Exchange:}
The exchange phase is the core of \vader, as shown in Fig.\ref{fig:vader_lifecycle}, where \MF\ and \MB\ perform exchange of the digital content.
Fig.~\ref{fig:sc_fsm} shows the messages used in the construction of \vader\ exchange protocol as described below. 
\newline

\begin{algorithm}[t]
  \caption{Offline Video Exchange \& \MB\ Ack}
  \label{apx:alg:video_exchange}
  \begin{flushleft}\ \ $t\ \triangleright$ Timeout \MB\ uses waiting for enc. content \\\end{flushleft}
  \begin{algorithmic}[1]
    \Function{\MF.SendContent}{\MCHAN, $C, reqid$}
      \State $k = Gen(\lambda)$\;
      \State $e_1,\dots,e_n = Enc^k(c_1),\dots,Enc^k(c_n)$\;
      \State $id_E = \{H(e_1),\dots,H(e_n)\}$\;
      \State \MCHANSEND($<E=<e_1,\dots,e_n>, id_E>$) to \MB\;
      \State $M_2 <id_E, \sigma_B>$ = \MCHANRECV() from \MB\;
      \State $\sigma_F = Sign(sk_F, id_E)$\;
      \State $M_3 = <id_E, \sigma_F, \sigma_B>$\;
      \State \MCHANSEND($M_3$) to \MB\;
      \State \MCHANST($M_2, M_3$)\;
    \EndFunction
  \end{algorithmic}
  \begin{algorithmic}[1]
    \Function{\MB.RecvContent} {\MCHAN, $reqid, t$}
      \State $<E, id_E> =$ \MCHANRECV() from \MF\;
      \State Save $E = <e_1, \dots, e_n>$\;
      \State $id'_E = \{H(e_1),\dots,H(e_n)\}$\;
      \ForEach{$i\ in\ id_E$}
        \If{$id'_E[i] \neq id_E[i]$}\;
          \State Request \MF\ for correct $e_i, id_E[i]$\;
          \State Start timer $t$; On Timeout:
          \State \ \ \textbf{if} correct chunk is received within $t$:\
          \State \ \ \ \ Update $e_i, id_E[i]$\;
          \State \ \ \textbf{else}\;
          \State \ \ \ \ $\triangleright$ Matching chunk \& hash is not received
          \State \ \ \ \ \MCHANCLOSE\;
        \EndIf
      \EndFor
      \State $\sigma_B = Sign(sk_B, id_E)$\;
      \State \MCHANSEND($M_2 <id_E, \sigma_B>$) to \MF\;
      \State $\triangleright$ $M_3$ is generated in SendContent
      \State $M_3 <id_E, \sigma_F, \sigma_B>$ = \MCHANRECV() from \MF\;
      \State \MCHANST($M_2, M_3$)\;
    \EndFunction
  \end{algorithmic}
\end{algorithm}

\begin{algorithm}
  \caption{Exchange}
  \label{apx:alg:exchange}
  \begin{flushleft}\ \ $t\ \triangleright$ Timeout \MB\ uses waiting for $k$ \\\end{flushleft}
  \begin{algorithmic}[1]
    \Function{\MB.SendIOU}{\MCHAN, $reqid, price_{id_C}, t$}
    \State $IOU = <I=pk_B,OU=pk_F,price_{id_C}>$\;
    \State $\sigma_B = Sign(sk_B, IOU)$\;
    \State \MCHANSEND$(M_4 <IOU, \sigma_B>) to$ \MF:\;
    \State \ \ \ \ Start timer $t$; On Timeout:
    \State \ \ \ \ \ \ \ \ \textbf{if} $k$ is not received \textbf{then}\;
    \State \ \ \ \ \ \ \ \ \ \ \ \ \ \MCHANCLOSE\;
    \State $M_5 <k, \sigma_F> =$ \MCHANRECV() from \MF\;
    \State \MB.$DecryptAndVerify(\text{\MCHAN}, reqid, k, \sigma_F)$\;
    \State \MCHANST($M_4, M_5$)
    \EndFunction
  \end{algorithmic}
  \begin{algorithmic}[1]
    \Function{\MF.SendKey} {\MCHAN, $reqid, k$}
    \State $M_4 <IOU, \sigma_B>$ = \MCHANRECV() from \MB\;
    \If{$Verify(pk_B, IOU, \sigma_B) \neq true$}
      \State \MCHANCLOSE\;
    \EndIf
    \If{$IOU.I \neq \text{\MB}\ or\ IOU.OU \neq \text{\MF}$}
      \State \MCHANCLOSE\;
    \EndIf
    \If{$IOU.price_{id_C} \neq price_{id_C}$}
      \State \MCHANCLOSE\;
    \EndIf
    \State $\sigma_F := Sign(sk_F, k)$
    \State \MCHANSEND($M_5 <k, \sigma_F>$) to \MB\;
    \State \MCHANST($M_4, M_5$)
    \EndFunction
  \end{algorithmic}
\end{algorithm}

\begin{algorithm}[t]
  \caption{Decrypt And Verify}
  \label{apx:alg:decrypt_verify}
  \begin{algorithmic}[1]
    \Function{\MB.DecryptAndVerify} {\MCHAN, $k, \sigma_F$}
    \State $c_1, \dots, c_n = Dec^k(e_1), \dots, Dec^k(e_n)$\;
    \State $id'_C = \{H(c_1),\dots,H(c_n)\}$\;
    \State $<cid, reqid, id_C, price_{id_C}> = \text{\MCHANLD}(M_1).m$
    \ForEach{$i\ in\ id_C$}
      \If{$id_C[i] \neq id'_C[i]$}\;
        \State $\triangleright$ Collect dispute evidence from channel
        \State $M = \text{\MCHANLD}(M_1, M_3, M_4, M_5)$\;
        \State \MCHANCLOSE
        \State $RaiseDispute(\text{\MB}, \text{\MF}, M, reqid, i, c_i)$\;
      \EndIf
    \EndFor
    \State $C = <c_1, \dots, c_n>$
    \State Everything OK. Continue Trade
    \EndFunction
  \end{algorithmic}
\end{algorithm}
\noindent\textbf{{Xchg.1} Exchange Agreement:}
The goal in this step is to ensure that both \MB\ and \MF\ mutually agree on the video to be
exchanged and the price to be paid by \MB\ to \MF. As depicted in Alg~\ref{apx:alg:trade_req}, \MB\ submits to
\MF\, the {\it Exchange Request message}, $M_0$, (m=<$cid$, $reqid$,
$id_C$, $price_{id_C}$>, $\sigma_B$), containing the channel id $cid$,
video identifier $id_C$, a \emph{randomly generated, strictly
monotonically increasing $reqid$}, as well as the amount
$price_{id_C}$ that \MB\ agrees to transfer on successful exchange. On
receiving $M_{0}$, \MF\ checks that the \emph{$reqid$ is indeed
monotonically increasing, to ensure $reqid$ is not reused (explained later
in penalization scheme)}, and the transfer price is agreeable. In return, \MF\
sends counter-signed message $M_1$ = <$M_0$,$\sigma_F$> to \MB,
committing to the terms.
\newline
\noindent\textbf{{Xchg.2} Out-of-Band Encrypted Video Transfer:}
Next, based on insight 1, as depicted in Alg~\ref{apx:alg:video_exchange} \MF\ randomly samples key $k$ and
sends encrypted version of the requested video 
along with signed hashes of each individual encrypted chunk
\newline
$id_{E}$=\{$H(E^k(c_{1}$ $\dots$ $c_{n}))$\} in $M_2$ to \MB.
\newline
On receiving encrypted video, \MB\ verifies if the digest matches with
the one sent by \MF\ in $M_2$. In case of a match, \MB\ sends counter
signed {\it Encrypted Video Acknowledgement} in $M_3$=<$M_2,\sigma_B$>. In case of
a digest mismatch, \MB\ requests \MF\ for retransmission of specific
chunks until either agreement is reached or failure after certain number of retries. 
In the next step, \MB\ and \MF\ exchange key and money in a trustworthy manner.
\newline
\noindent \textbf{{Xchg.3} Money-Key Exchange:}
In this step, as shown in Alg~\ref{apx:alg:exchange}, \MB\ \textbf{1)} sends an {\it optimistic IOU}
message $M_4$ to \MF, and \textbf{2)} starts a timer $t$ within which
\MF\ needs to send the key $k$. In case of timeout, \MB\ closes the
channel and initiates dispute resolution. On receiving the IOU,
\MF\ releases the key along with digest in $M_5$.
\newline
\noindent \textbf{$\bullet$ Phase 3: Finalization:}
In this phase, \MB\ verifies whether it received the right content or
raises a dispute.
\newline
\noindent\textbf{{Final.1} Verification:}
As shown in Alg~\ref{apx:alg:decrypt_verify}, \MB\ decrypts the video using $k$ and verifies that 
$\{H($$Dec^k(c_{1}$ $\dots$ $c_{n}$)\} matches $id_C$ uploaded by \MO. In case
of match, \MB\ has received desired content and the same channel can be re-used for future exchanges.
However, in case of a mismatch, i.e. \MF\ sent wrong content, \MB\ closes the channel and registers a dispute with the {\it Dispute Resolve} smart-contract by submitting
$M=<M_1,M_3,M_4, M_5>$ as evidence along with at least one of the mismatched chunks, $c_i$.
\newline
\noindent \textbf{{Final.2} Channel Balance Update:}
At the end of a successful exchange, each party locally updates the
channel balance of the counterparty and decides if the channel has
sufficient balance to continue or must be closed.
\newline
\indent As shown in the Fig.~\ref{fig:vader_lifecycle}, in both phase 2 \& 3, \MB\ \& \MF\ {\bf do not}
interact with blockchain and carry out opportunistic exchanges (based on insight 2).
\newline
\begin{algorithm}[t]
  \caption{Dispute Resolve Smart Contract}
  \label{alg:dispute_resolve}
  \begin{flushleft}$M=<M_1,M_3,M_4,M_5>\ \triangleright$ Dispute evidence\end{flushleft}
  \begin{algorithmic}[1]
    \Function{RaiseDispute} {\MB, \MF, $M, reqid, i, c_i$}
      \State $<id_E, \sigma_F, \sigma_B> \leftarrow M.M_3$
      \State $<IOU, \sigma_B>,<k, \sigma_F> \leftarrow M.M_4,\ M.M_5$
      \If{$k$ is $null$ or $H$($E(k,c_i))$ $\neq$ $id_E(i)$}
        \State Start timer $\tau$; On Timeout:
        \State \textbf{if} (No $k$ from \MF\ within $\tau$) OR\;
	  \State \ \ (\MF\ gives $k$ \textbf{but} $H$($E(k,c_i))$ $\neq$ $id_E(i)$)
        \State \ \ \ \ \ \textbf{goto} ReturnMoneyTo{\MB}\;
      \EndIf
      \State $<cid, reqid, id_C, price_{id_C}> \leftarrow M.M_1.M_0$
      \If{$H(c_i) = id_C(i)$}
	  \State $\triangleright$ \MB\ Cheated, \textbf{No Loss to} \MF\
        \State \Return
      \Else\ $\triangleright$ \MF\ Cheated.
      \EndIf
      \newline
      ReturnMoneyTo{\MB}:
          \State \ \ \ \MF.$escrowBalance$ -= $price_{id_C}$\;
          \State \ \ \ \MB.$escrowBalance$ += $price_{id_C}$\;
  \EndFunction
  \end{algorithmic}
\end{algorithm}

\noindent \textbf{$\bullet$ Phase 4: Settlement:}
At the end of multiple exchanges, \MF\ and \MB\ submit all
their offchain state to blockchain smart contracts ({\it trusted arbitrators}) to settle channel
balance, resolve disputes and automatically pay \MO\ based on the number of successful exchanges.
\newline
\noindent \textbf{{Settle.1} Channel Closure and Settlement:}
Akin to state-channels, \vader\ channel closing semantics guarantee that once a party
closes the channel, the other party has up to time $\tau$ to submit its
set of offchain signed messages as evidence to the {\it channel settlement} smart
contract. After time $\tau$, the smart contract verifies the validity
of each offchain message sequence and first settles the successful
exchanges by transferring money worth $\sum_{i}price_{id_C}$ to \MF\
and $\sum_{i}(amt_{O}\%)$ to \MO\ from channel escrow.
Disputes are settled as described next.

\noindent\textbf{{Settle.2} Dispute Resolution:} 
In case of a dispute raised by \MB\, the {\it Dispute Resolve} Contract
Alg.~\ref{alg:dispute_resolve} performs two steps to identify the
faulty participant as described below
\textbf{1)} In line 4, it encrypts the disputed chunk with key $k$ (if present) and checks if
the hash of the encrypted chunk matches the one agreed by \MB\ in $M_3$.
In case of {\it key not released} dispute by \MB, \MF\ has time $\tau$ within which to
submit $k$. 
\textbf{2)} In line 10, it computes hash of the chunk and checks if it matches with the hash registered by \MO\ in \emph{Init.1}.
A {\bf faulty facilitator} will fail step (2) if chunk is indeed different from the one registered by \MO; otherwise, the request is discarded because of {\bf faulty buyer} behavior.
Once the faulty party is detected, channel escrow balance is appropriately transferred to the honest party. 

At the end of settlement and dispute resolution, offchain
channel state is transferred on-chain, and application progresses, disputes continuing
directly on blockchain with each party submitting messages directly to
blockchain instead of each other.
\newline
We note that \vader\ automatically pays out royalty to \MO\ based on the
successful exchanges submitted to blockchain during settlement
phase. Therefore, \vader\ guarantees \MO\ fairness as long as \MB\ and
\MF\ {\it honestly} report their offchain messages on chain and do not
collude with each other. We next describe how \vader\ handles
collusion between a malicious \MB\ and \MF\ and still guarantees \MO\ fairness.  


\begin{algorithm}[t]
  \caption{Penalizer Smart Contract}
  \label{alg:penailzer}
  \begin{flushleft}\ \ $bounty$: penalty amount\end{flushleft}
  \begin{flushleft}\ \ $cid$: channel id for <\MB\,\MF>\end{flushleft}
  \begin{algorithmic}[1]
    \Function{\MB.SubmitClaim} {\MB, \MF, $cid, M_1, M_1'$}
    \State $<cid, reqid, id_C, price_{id_C}, \sigma_B, \sigma_F> \leftarrow M_1$
    \State $<cid', reqid', id'_C, price'_{id_C}, \sigma'_B, \sigma'_F> \leftarrow M'_1$
    \If{$VerifySigns(\text{\MB}, \text{\MF}, [M_1, M'_1]) \neq true$}
      \State $\triangleright$ Message signatures Invalid.
      \State \Return\;
    \EndIf
    \State \textbf{if} (<$reqid$,$cid$> $=$ <$reqid'$,$cid'$>)\ AND 
    \State \ \ \ \ $(id_C \neq id'_C)$\ \textbf{then}
    \State \ \ $\triangleright$ \MF\--\MB\ collusion attack detected
    \State \ \ $\triangleright$ Penalize \MF\ and reward \MB\
    \State \ \ \MF$.escrowBalance$ -= $bounty$
    \State \ \ \MB$.escrowBalance$ += $bounty$
    \EndFunction
  \end{algorithmic}
\end{algorithm}

\subsection{Preventing \MF\--\MB\ Collusion}
\label{sec:owner_rights}
The offchain execution of \vader\ between \MB\ and \MF\ makes \MO\ 
vulnerable to a collusion attack~\footnote{The problem does not occur
in case one of \MB\ or \MF\ are honest, as the adherence to protocol
by any one party will force the other party to act honestly or loose
out.} where in \MB\ and \MF\ submit an altered set of offchain message
exchanges during settlement to deny \MO\ of its fair share. We explain how
such collusion can be undertaken next, followed by the description of the technique used to prevent such attacks.

\noindent \textbf{Offchain Alternate Message Construction}:
Consider the case where \MB\ and \MF\ have completed a
successful exchange of video $id_C$ owned by \MO. Note that at
this point, \MF\ has already received $IOU$ and \MB\ the desired
file. Therefore, both parties are incentivized to collude to maximize
profit by constructing a new set of message exchanges simulating
\textbf{1)} \MB\ requesting video $id_C'$ owned by a sybil entity (\MO') controlled by \MF\
benefitting \MF\ , \textbf{2)} \MF\ agreeing for price $price_{id_C'}$ much lesser than
$price_{id_C}$ benefiting \MB. As part of alternate message construction, \MB\ creates a
new Exchange Request Message $M_{0}'$, (m=<$cid$, $reqid$, $id_{C'}$, $price_{id_C'}$>,$\sigma_B$) with video id
${id_{C'}}$. Note that a {\it rational} \MB\ will reuse the <$cid$,$reqid$>
pair, since creating either a new $cid'$ or $reqid'$ makes \MB\
vulnerable to \MF\ submitting it as legitimate evidence of exchange to
blockchain, and withdrawing \MBs\ money.
We leverage the above nuanced message construction to detect and penalize collusion.

\noindent\textbf{Penalizer Smart Contract}:
Specifically, following insight 3, we introduce another smart contract, {\it Penalizer} Alg~\ref{alg:penailzer},
which will pay \MB\ a large penalty from \MF's
funds (deposited during channel funding in \emph{Init.2}), if it can submit a pair of $M_1,M'_1$ messages, in line 2 \& 3, 
having same \emph{reqid} and \emph{cid} but different content id's $id_C,id'_C$. The penalization scheme introduces an element of distrust between \MB\ and \MF\ and prevents collusion, forcing them to
honestly report their offchain exchanges during settlement.
In summary, our penalty scheme ensures that in a realistic setting,
where most buyers are not controlled by \MF, the \MF\ is disincentivized to
collude with \MB, for fear of paying a heavy penalty. Consequently our
protocol guarantees fairness to all three parties viz. \MO, \MF\ and \MB.
\newline
\noindent\textbf{Limitations} We note that the offchain nature of
\vader\ entails a buyer (or its delegatee~\cite{pisa}) to
continuously remain {\it online}, and maintain {\it significant local state}, as
well as lock up {\it sufficient liquidity} in channel till
settlement. Consequently, \vader\ may not be appropriate for light
weight clients that are ephemeral, lack local state or cannot afford
to lock liquidity for long time, and depend solely on blockchain for
tamper-proof logging and availability. We evaluate this model in Sec.~\ref{sec:implementation}.

\noindent\textbf{Extension to multiple \MO, \MF, \MB\ :}
In reality multiple parties can assume any of the roles in a single exchange (say multiple \MO\ of same content). Note that
our protocols will work for even in such settings as long as all parties {\it do no completely trust each other}. 
This is due to the fact that any new entities participation will either be essential to the exchange and in which case, their concerns will be similar to that of an active party, whose rights are protected by \vader\ protocol. Alternatively, the party might not be essential for completing the exchange, in which case, their concerns will be similar to that of passive party which are also safeguarded by \vader\ protocol.

\subsection{Security Analysis of \vader\ }
\label{sec:security}
In this section we show that \vader\ is secure against Royalty Manipulation,
Content Mismatch and Content Stealing attacks (\textit{Atk.1,2,3})  defined in Sec.~\ref{sec:setting}.
We reiterate that as mentioned in Sec.~\ref{sec:setting} we do not address the orthogonal issue of content piracy.
\begin{theorem}
Under the assumptions that,
a) the digital signature scheme is unforgeable,
b) the cryptographic hash function is collision-resistant and
c) under honest majority, blockchain attributes (smart-contract execution and ledger) are tamper-resistant,
we show that \vader\ protects fairness of \textbf{honest and rational parties} against {\it Atk.1, 2 \& 3}.
\end{theorem}
\noindent \textbf{$\bullet$ Atk.1:}
\vader\ ensures that in presence of at least honest \MF\ or \MB, all exchanges are recorded onto the blockchain during \textbf{channel close} either successfully or as a dispute. In case both \MF\ \& \MB\ are malicious, \vader\ introduces distrust to prevents collusion between them through incentive techniques mentioned in Sec.\ref{sec:owner_rights}, forcing them to record correct state on the blockchain.
Further, in \vader\ \MO's royalty is calculated inside the blockchain trusted arbitrator (smart contract).
So under the assumption (c) any royalty manipulation attack will not be feasible given honest majority of blockchain participants. 

\noindent \textbf{$\bullet$ Atk.2 \& 3:} We analyse {\it Atk.2, 3} in individual malicious parties and collusion cases as follows:

\noindent \textbf{$\rightarrow$ Malicious \MO:}
A malicious \MO\ may target a content mismatch attack ({\it Atk.2}) by sending mismatching content ($C,id_C$)  in
{\it Init.1}, but for an honest \MF\ to accept mismatching content and hashes, \MO\ needs to find a collision in the hash function which is assumed impossible, making this attack infeasible.
Content stealing ({\it Atk.3}) is useless for \MO\ as it owns the very content.
\newline
\noindent \textbf{$\rightarrow$ Malicious \MF:}
For a malicious \MF\ to successfully launch a content mismatch ({\it Atk.2}), it has to send a content $C'$ instead of $C$ as requested by \MB. An honest \MB\ will not accept that 
unless \MF\ manages to find a collision in $H$ such that $H(C')=H(C)$. Hence, under collision-resistant hash function assumption, {\it Atk.2} does not occur. Content stealing ({\it Atk.3}) is meaningless for \MF\ and will not occur as it arguably owns the very content.
\newline
\noindent \textbf{$\rightarrow$ Malicious \MB:}
A malicious \MB\ can try to mount content mismatch attack {\it Atk.2} by raising a fraudulent complaint
against \MF\ after receiving the correct content. However, since \MB\ and \MF\ exchange a co-signed message $M_3$ agreeing to ($E^k(C),id_E$),
\MB\ cannot generate \MFs\ sign on a ($E^k(C'),id_E'$) due to the unforgeability of signatures assumption. Similarly, it cannot raise a complaint with {\it Dispute Resolve} without submitting a mismatching chunk which when encrypted under $k$ leads to the same hash, without finding a collision in the 
hash function, which is assumed impossible.
\newline
\indent A malicious \MB\ cannot steal content({\it Atk.3}) from \MF\ as as an honest 
\MF\ will not transfer the decryption key $k$ without payment in message $M_4$ and {\it Dispute Resolve} contract 
ensures that \MB\ provides $IOU$ to \MF\ before handing over the key $k$.
\newline
\noindent \textbf{$\rightarrow$ \MO\--\MF\ Collusion:}
A malicious \MO\ cannot help \MF\ with content mismatch attack ({\it Atk.2}) to cheat \MB\ by providing wrong content
as an honest \MB\ requests for a particular content in the Xchg.1 step.
Since \MO\ has no control over this step, this attack reduces to a malicious \MF\ case, which we already explained the 
protocol to be secure against. Also both \MO\ and \MF\ arguably own the content, hence obviating the need for content 
stealing attack ({\it Atk.3}).
\newline
\noindent \textbf{$\rightarrow$ \MB\--\MO\ Collusion:}
Content mismatch ({\it Atk.2}) \& stealing ({\it Atk.3}) attacks in this case would be \MB\ and \MO\ trying
to exchange content through \vader\ but denying payment to \MF.
To this end, \MO\ may try to submit wrong ($C$,$id_C$) in Init.1 step, in 
order for \MB\ to raise a dispute. An honest \MF\ will not countersign a mismatching ($C,id_C$), unless \MO\ finds a 
collision in $H$ which by assumption is impossible. Apart from Init.1, \MO\ does not participate in 
the remaining protocol, reducing these attacks to the malicious \MB\ case, which the protocol is secure against.
\newline
\indent Note that the case of mutually known \MO\ and \MB\ carrying out entire exchange among themselves without involving \MF, is orthogonal to the \vader\ setting wherein \MB\ and \MO\ do not know each other. 
\newline
\noindent \textbf{$\rightarrow$ \MB\--\MF\ Collusion:}
Rational \MB\ and \MF\ will not let a content mismatch attack ({\it Atk.2}) and content stealing attack ({\it Atk.3}) to occur respectively as they are hurt by the same. We ignore the case where a \MF\ (or any fake \MO'\ ) sells the content to a \MB\ outside our ecosystem as that is a case of content piracy.
\newline
\indent Under all the possible cases (corruption and collusion), \vader\ is secure against {\it Atk.1, 2\ and\ 3}, thus completing the proof.
We briefly highlight the reason for \vader\ to be secure even with multiple \MF, \MO\ or \MB\ in Sec.~\ref{sec:owner_rights}.

\section{Implementation}
\label{sec:implementation}
We compare the performance of \vader\ against the following two baselines
\newline
\noindent \textbf{VANILLA}: We implemented \vanilla\ to emulate traditional HTTPS based video delivery as described in RFC~\cite{rfc_http_streaming}.
\vanilla\ protocol {\bf does not} use blockchain enabling us to benchmark performance overhead of \vader\ over state-of-art mechanisms.
\newline
\noindent\textbf{Blockchain Mediated Exchange (BME)}: To understand the
benefits of batching protocol messages, we implement \fullon\ protocol
wherein \MF\ and \MB\ progress application state directly on blockchain instead of directly sending to each other as in \vader.
In \fullon, akin to \vader, video is transferred offchain in encrypted format,
while each exchange is directly settled onchain before starting the next.
We note that in \fullon\ parties need not open \& close a channel, and some messages can be batched while committing to blockchain, requiring only 3 commits (i.e. <$M_0$,$M_1$>, <$M_2$,$M_3$,$M_4$> \& <$M_5$>) in {\it happens before order}. We incorporate the above optimizations in our implementation of \fullon.
\newline
\noindent{\bf Application Prototype:} We implement \MF\ as a Django webapp (868 SLOC), \MO\ (256 SLOC) \& \MB\ apps for \vader\ 
(699 SLOC), \vanilla\ (348 SLOC) and \fullon\ (617 SLOC), as Python applications.
\newline
\noindent{\bf Smart Contracts:}
We implement all \vader\ and \fullon\ functionalities as chaincodes (Smart Contracts) on top of Hyperledger Fabric~\cite{hyperledger_fabric} v1.2 (Fabric)~\cite{hyperledger_fabric_paper}. 
We implement chaincode  utilities for time estimation based on block height and conditional
(time and condition locked) escrow accounts (no native crypto-currency in Fabric v1.2). 
\newline
\noindent{\bf Maliciousness:} We emulate a malicious \MF\ as one that sends non-matching chunk hashes to \MB. Similarly,
a malicious \MB\ is emulated by raising a dispute with the {\it Dispute Resolve} smart contract even after receiving 
the correct video from \MF. 

\section{Evaluation}
\label{sec:evaluation}

\begin{figure*}[t!]
    \begin{minipage}[t]{0.66\textwidth}
        \begin{subfigure}[t]{0.4\textwidth}
            \hspace*{-0.7cm}
            \includegraphics[height=1.5in]{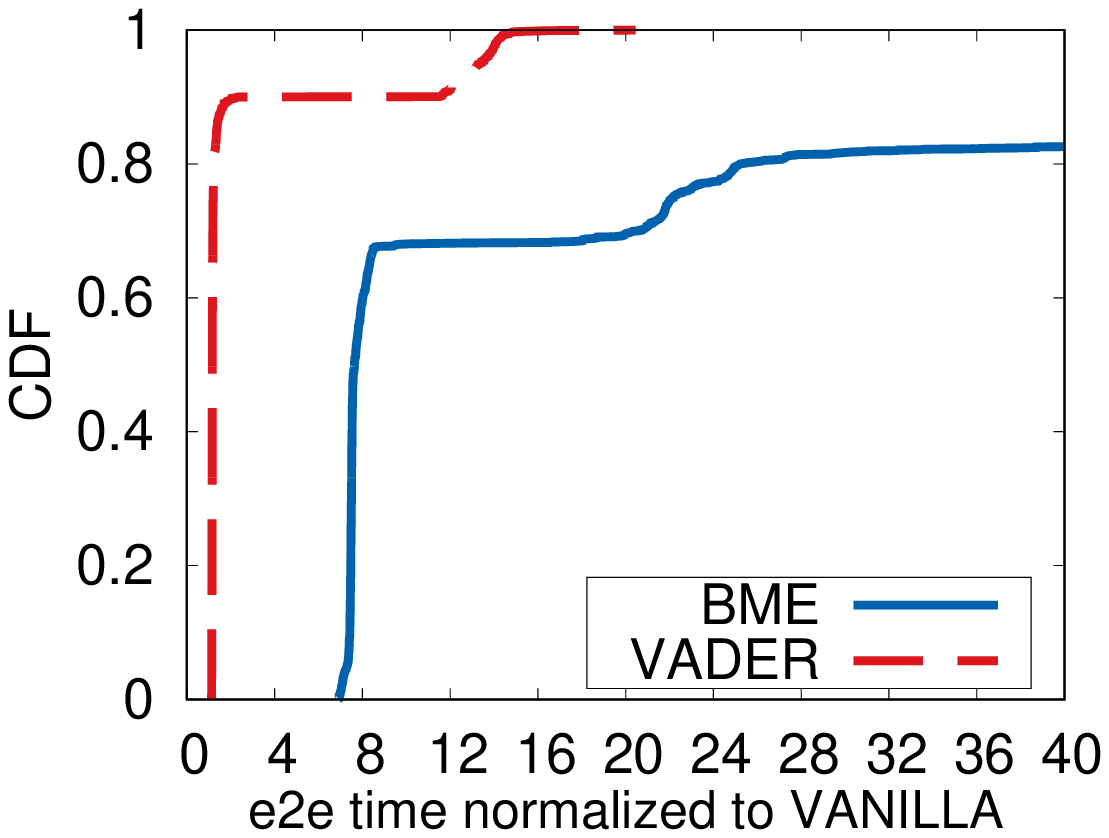}
            \caption{CDN Topology}
            \label{fig:intradc_sc_and_fullon_vs_vanilla_CDFs_20M_median}
        \end{subfigure}
        ~
        \begin{subfigure}[t]{0.5\textwidth}
            \includegraphics[height=1.5in]{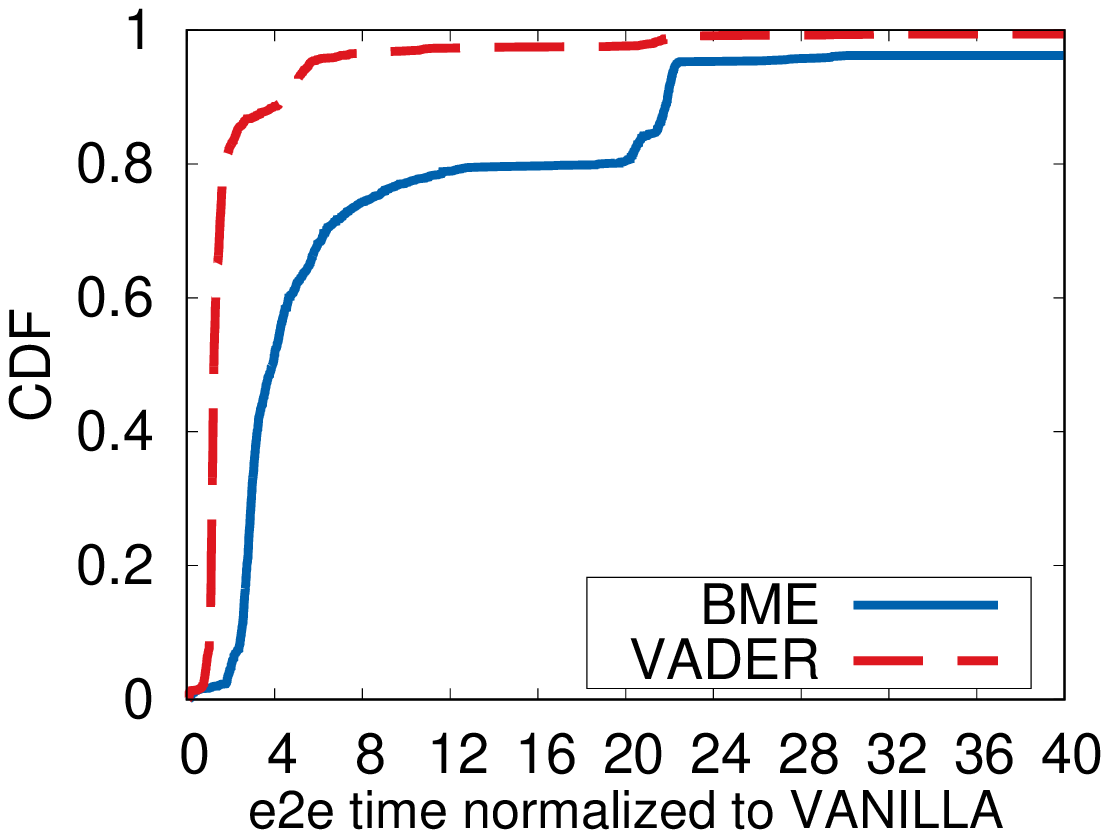}
            \captionsetup{margin={1.8cm}}
            \caption{Random Topology}
            \label{fig:interdc_sc_and_fullon_vs_vanilla_CDFs_5M_median}
        \end{subfigure}
        \vspace*{-0.2cm}
        \captionsetup{justification=centering}
        \caption{E2E time of \vader\ and \fullon\ normalized to \vanilla\ with
                \#\MB\ 500, \#\MF\ 10, 10\% maliciousness exchanging random \#files 10-250}
        \label{fig:CDFs_normalized_VANILLA}
    \end{minipage}
    ~
    \begin{minipage}[t]{0.33\textwidth}
        \includegraphics[height=1.5in]{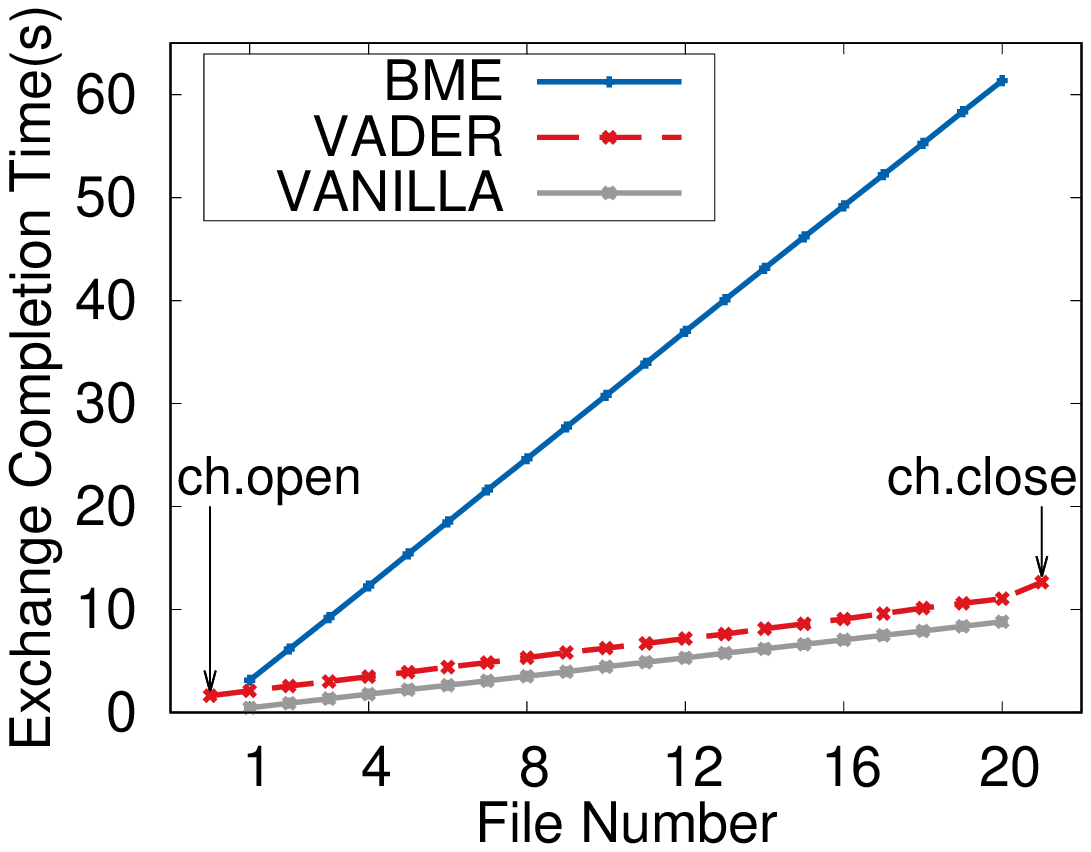}
        \caption{File exchange timeline for an honest \MB\ exchanging 20 files, 20MB each under CDN topology}
        \label{fig:intradc_file_download_timeline}
    \end{minipage}
\end{figure*}

In our evaluation, we answer the following,
\textbf{1)} the performance overhead of \vader\ compared to baseline protocols,
\textbf{2)} the amortization benefits of \vader,
\textbf{3)} the effect of maliciousness on \vader\ performance,
\textbf{4)} the sensitivity of \vader\ to the underlying blockchain platform.

\noindent\textbf{$\bullet$ Experimental Setup:}
We run our experiments on 91 VMs (Ubuntu, 2.1GHz 16 CPU, 32GB, SSD)
in Softlayer Cloud~\cite{ibm_softlayer}, across five geo-distributed
data centers spanning four continents. Based on benchmarking, average
latency was estimated at 0.4ms (and throughput 5Gbps) for intra-DC and
varied 21.8-337ms (throughput 460-35.6Mbps) for inter-DC network.
All experiments were conducted with a Fabric network consisting of 10 blockchain peers
(one per \MF\ reflecting a network run by the \MFs) running the default
ordering service with out of the box configuration parameters~\cite{hyperledger_fabric_github}.
\newline
\noindent\textbf{$\bullet$ Performance Metrics:}
We use the following metrics for quantifying performance
overhead of the three schemes:
\textbf{1) end-to-end (e2e) time} measures time elapsed from the instant \MB\ requests a file till it's last chunk is ready for consumption (downloaded and decrypted) is our primary metric;
\textbf{2) component contributions} measured as the fraction of time
spent in each sub~component obtained
by dividing e2e time into three components--
\textbf{a) Protocol Time} is the total time spent in blockchain
interactions and message exchanges between the parties;
\textbf{b) Transfer Time} is the time taken for transferring encrypted chunks;
\textbf{c) Verify Time} is the time spent by \MB\ in decrypting
and verifying hash of the exchanged file. In the presence of maliciousness,
this sub-component includes the time spent in interacting with the dispute resolution smart-contract;
\newline
\noindent \textbf{$\bullet$ Experimental Methodology:} 
We benchmark the performance of \vanilla, \vader\ \& \fullon\
under different realistic conditions
by varying the underlying network topology, load on \MF\ and overall
maliciousness in the system.
Unless stated otherwise, we run our experiments for five iterations and report average results.

\subsection{Macro Benchmarks} \label{subsec:macro_bench}
In this section we benchmark the performance overhead of all three schemes 
under realistic settings. We run 500
\MBs\ in 5~DCs (100 \MBs\ per DC, equally load-balanced among 5 VMs).
The \MBs\ are configured to exchange a 20MB file in chunks of 512KB~\footnote{Empirically
determined to be optimal, results omitted for sake of brevity.}, random number
of times (10 to 250, in increments of 5). We run 10 \MFs\, in 10 VMs equally distributed across the 5~DCs. We configure 10\% of the
\MFs\ to be malicious (as described in Sec.~\ref{sec:implementation}).


\noindent \textbf{\textit{5.1.1} CDN topology:}
We emulate a CDN like hierarchical topology implemented by real
world facilitators~\cite{YoutubeCDN,UnreelingNetflix} where a \MB\
exchanges content from the nearest (same DC) \MF\ and measure e2e 
time for all the three schemes.

\noindent{\textbf{Performance Overhead: Non-Malicious:}}
In Fig.~\ref{fig:intradc_sc_and_fullon_vs_vanilla_CDFs_20M_median},
we plot the CDF of (median) e2e time for \vader\ and
\fullon\ \MBs, normalized to \vanilla. \vader\ adds only minimal overhead
to \vanilla\ ranging from min. 12\% to 16\% at the median, making it a
practically deployable system for providing fairness at scale. Notably
80\% of the \vader\ \MBs\ have an overhead of less than 23\%.
In contrast, \fullon\ has an overhead ranging from min. 690\% to 764\% at the median.
This is due to the fact that \vader\ interacts with blockchain only
twice (channel open and close) over multiple successful exchanges,
while \fullon\ interacts with blockchain thrice for each exchange.

\emph{In summary, \vader\ is able to amortize blockchain interaction
  time over multiple exchanges leading to minimal performance overhead compared to \vanilla\ (16\%) whereas, lack of amortization
leads to significant degradation in \fullon\ (764\%).}

\noindent \textbf{Performance overhead: Malicious:}
We also observe that maliciousness (10\% of \MFs) causes severe
performance degradation shown by a lot worse
performance (notch at 90\%) for both \vader\ and \fullon.
\vader\ adds an overhead of atleast 1160\% while \fullon\ adds 9400\%
respectively for clients of malicious \MFs.
The performance degradation can be attributed to the overhead imposed by 
onchain execution of \textit{Dispute Resolve} contract involving sign verification and
waiting for settlement timeouts.
as described in Alg.~\ref{alg:dispute_resolve}.

\begin{table*}[th!]
    \begin{adjustbox}{width=\textwidth}
    \centering
    \begin{tabular}{|l|c|c|c|c|c|c|c|c|}
    \hline
    Blockchain & Bitcoin & Ethereum & Litecoin & Siacoin & Monero & Zcash & Peercoin & Dogecoin \\
    \hline
    Consensus & PoW & PoW & PoW & PoW & PoW & PoW & PoW \& PoS & Pow \\
    \hline
    Block Gen. Time(s) & 545.52 & 14.58 & 149.82 & 600.00 & 121.56 & 150.00 & 484.38 & 62.52 \\
    \hline
    \vader\ (s) & 5.88 & 0.57 & 1.92 & 6.42 & 1.64 & 1.92 & 5.26 & 1.05 \\
    \hline
    \fullon\ (s) & 1637.04 & 44.22 & 449.95 & 1800.48 & 365.17 & 450.48 & 1453.63 & 188.04 \\
    \hline  
    \end{tabular}
\end{adjustbox}
   \caption{\vader\ \& \fullon\ average e2e expected latencies (per 20MB file) for public blockchain networks while batching 200, 20MB file exchanges in a session. Block generation time for Siacoin is from~\cite{coingecko_siacoin}, for rest, 1/1/2019 from~\cite{bitinfocharts}. PoW = Proof Of Work, PoS = Proof of Stake}
    \label{tab:public_blockchains_table}
\end{table*}

\noindent \textbf{\textit{5.1.2} Random topology:}
Next we evaluate the performance of \vader\ under a different network
topology viz. random. We repeat the same experiment above, but allowing \MBs\ to
exchange content from a random \MF\ in any DC this time~\footnote{
Due to inter-dc network limitations and also to reduce the financial cost
of the experiment, we restrict our file size to 5MB, keeping number of exchanges
same (10 to 250). (To validate results, we have manually run
experiments with 20MB files and found similar trends as reported).
}.

Fig.~\ref{fig:interdc_sc_and_fullon_vs_vanilla_CDFs_5M_median}
shows that compared to \vanilla, \vader\ has a minimal overhead of
23\% at the median compared to \fullon's 391\%.
Interestingly, \fullon\ has lesser overhead in this scenario compared
to the CDN scenario. This is due to the inter-dc network conditions that make network transfer times relatively worse for a large number
of \MBs\ across all three schemes. Consequently, overall network
transfer time increases while the blockchain overhead
remains the same making \fullon\ incur relatively lesser overhead compared to CDN topology.

\emph{Note that, \vader\ maintains its minimal overhead (23\%)
  compared to \vanilla\ even under adverse network conditions making it
  deployable over a variety of network topologies.
}

\noindent \textbf{\textit{5.1.3} Analysis of a single CDN experiment:}
To get a better understanding of when a file is available for
viewing by a \MB\ (startup latency), we randomly select a single \MB\ from the CDN
experiment.
and depict the timeline of all file exchanges 
(20MB, 20 times), for all three schemes in Fig.~\ref{fig:intradc_file_download_timeline}.
We note that, barring the one time channel open delay (1.61s), \vader\ adds only a minimal delay compared to \vanilla\ for each file, thereby not
adversely affecting user experience. On the other
hand, \fullon\ adds significant delay per file due to its three commit
blockchain overhead added to each file.
{\it We also observe that, barring
channel open and close overheads (3.2s), \vader\ adds only 7\% delay
over vanilla, making it a viable system.}


\noindent \textbf{\textit{5.1.4} Estimated performance on Public Blockchain Networks:}
We evaluate the sensitivity of \vader\ to underlying blockchain platform by
estimating it's performance on various public blockchains listed
in Tab.~\ref{tab:public_blockchains_table}. For this study, we calculate the median e2e time of a single file for \vader\ \& \fullon\ and
isolate it into two components viz. `blockchain protocol' time and
`miscellaneous' time involving network transfer and crypto operations. We estimate blockchain protocol time for \vader\ by dividing twice the block generation time of the underlying blockchain (one each for channel open-close) by the
number of file exchanges. For \fullon\, we calculate protocol time as thrice the block generation time (corresponding to the three blockchain commits).
Finally we add the miscellaneous overhead for both protocols to get the projected time taken by each.

We observe that even in the case of public blockchains,
\vader's amortization benefits drastically outperform \fullon.
In the case of blockchain
networks with high block generation time (such as Bitcoin)
\vader\ is able to achieve 27.21Mbps, making it practical even on
public blockchains, while \fullon\ throttles down to 0.01Mbps.
\emph{On the other hand, in the case of a blockchain like Ethererum 
with lower block generation time, \vader\ can achieve nearly 280Mbps,
which is comparable to \vanilla's 384Mbps.}
\subsection{Micro benchmarks}
\label{subsec:micro_bench}
In this section, we validate our design choices by benchmarking
the performance of various system sub components.
We allocate 10 VMs running a single \MF\ each, and run 200
\MBs\ in 20VMs (20 buyers per vm). We configure \MBs\ to exchange a 20MB file in chunks
of 512KB, 50 times randomly from any \MF. Default maliciousness is set to 0\% (unless specified otherwise).

\begin{figure*}[t]
    \centering
    ~
    ~
    \begin{minipage}[t]{0.35\textwidth}
        \hspace*{-0.9cm}
        \includegraphics[height=1.5in]{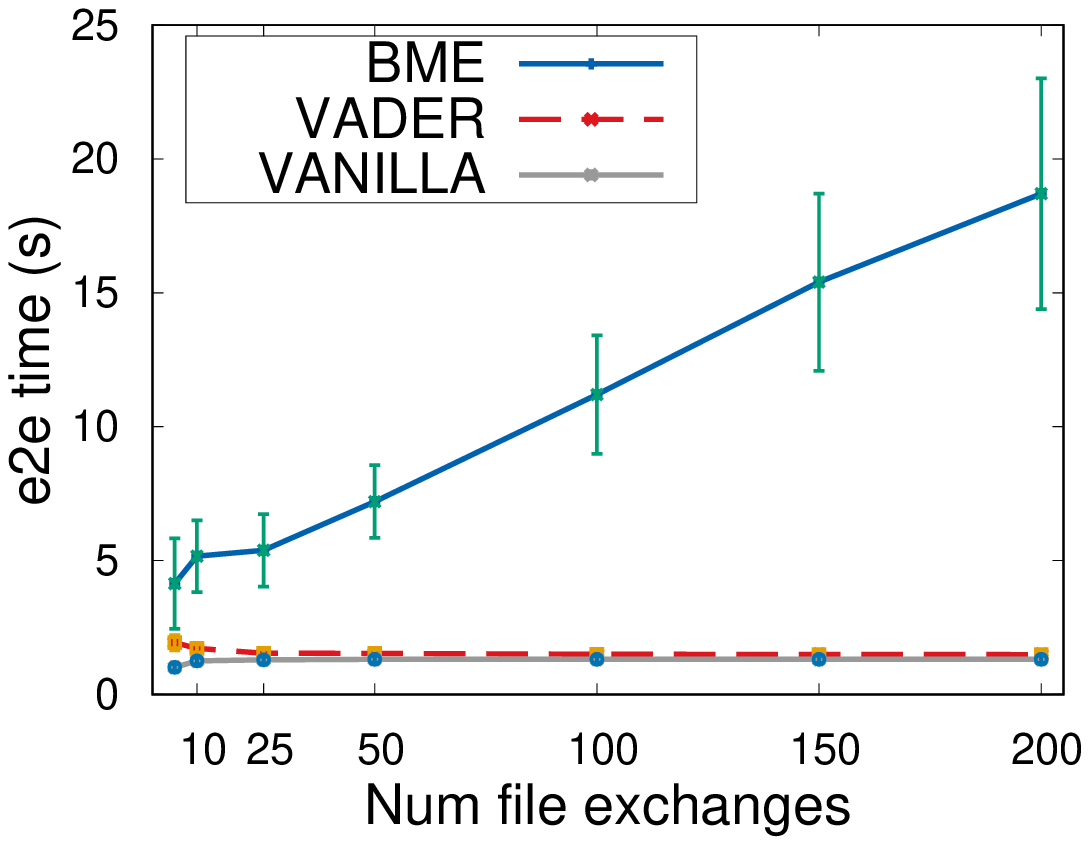}
        \caption{E2E time amortization w. increasing no. of file exchanges. (filesize 20MB)}
        \captionsetup{margin={1.2cm}}
        \captionsetup{justification=centering}
        \label{fig:micro_numiterations_vs_e2e}
    \end{minipage}
    ~
    ~
    \begin{minipage}[t]{0.3\textwidth}
        \hspace*{-0.6cm}
        \includegraphics[height=1.5in]{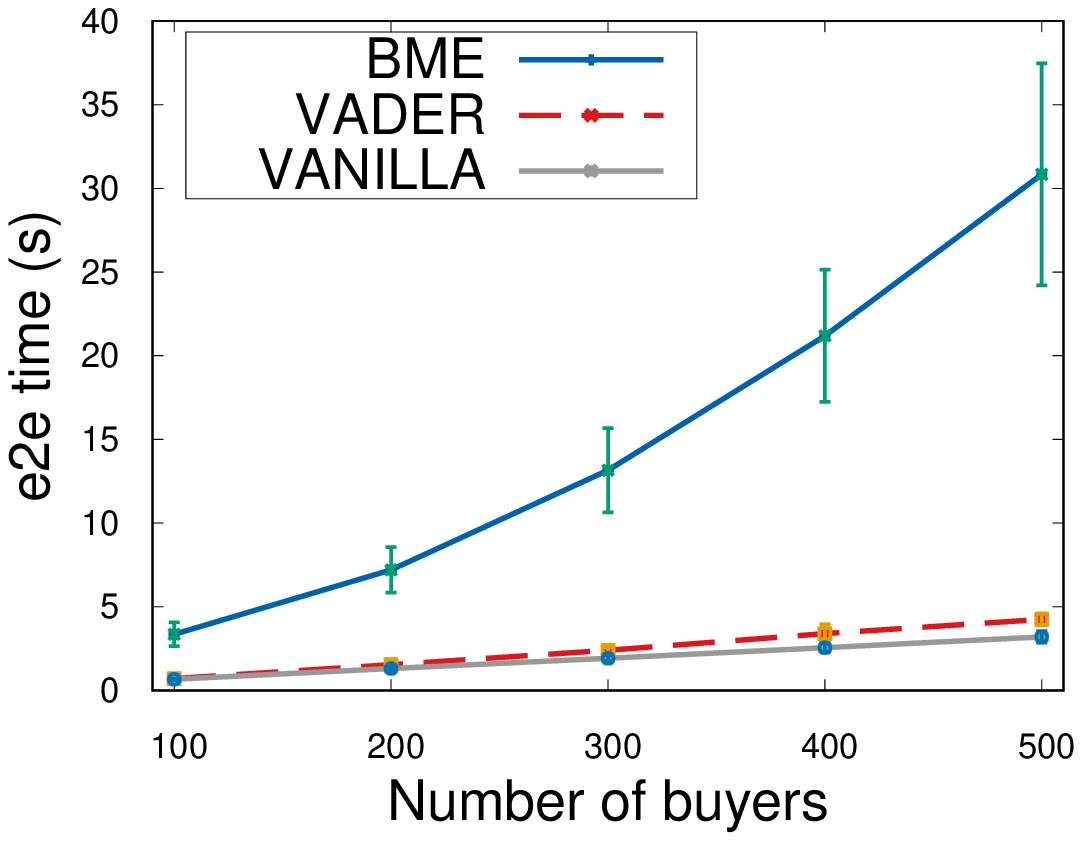}
        \captionsetup{justification=centering}
        \caption{E2E time vs increasing \#\MBs\ (20MB, 50 times)}
        \label{fig:micro_numbuyers_vs_e2e}
    \end{minipage}
    ~
    ~
    \begin{minipage}[t]{0.3\textwidth}
        \includegraphics[height=1.5in]{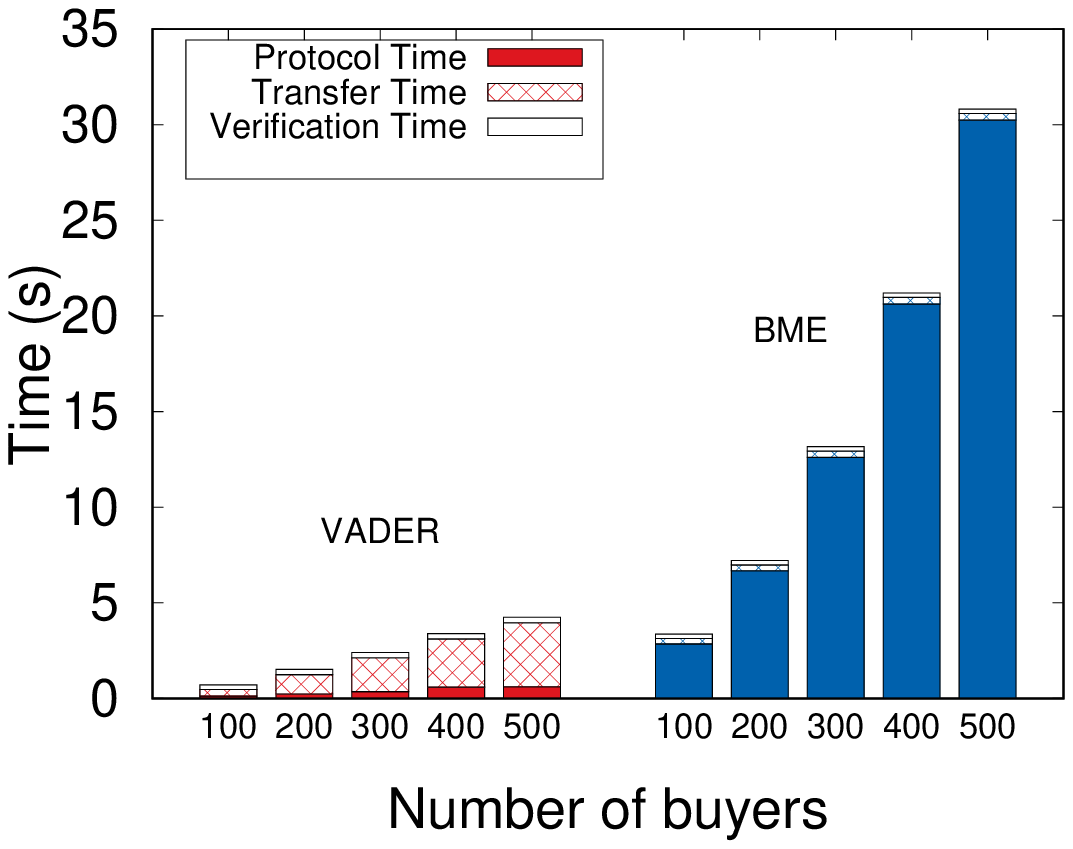}
        \captionsetup{justification=centering}
        \caption{Per component e2e time contribution vs \#\MBs\ (20MB, 50 times)}
        \label{fig:micro_component_vs_buyerload}
    \end{minipage}
\end{figure*}

\noindent \textbf{\textit{5.2.1} e2e time vs \#files:}
Fig.~\ref{fig:micro_numiterations_vs_e2e} shows the effect of increasing
\#files (5 to 200) on average e2e time per file for a
constant file size of 20MB. We observe that \vader\ overhead compared to \vanilla\ keeps
decreasing with increasing no. of files, from 90\%(5 files) to
13\%(200 files). 
Clearly, \vader\ is able to amortize blockchain overhead over multiple
files. On the other hand, note that the e2e time for \fullon\ increases rapidly
with the no. of files, from 4.14s(5) to 18.7s(200) since \fullon\ bottlenecks by hitting the blockchain for each file.


\noindent \textbf{\textit{5.2.2} e2e time vs \#buyers:}
Next we study the effect of loading \MF\ by increasing no. of \MBs\ from
10 to 50 per \MF, keeping filesize (20MB) and \#files (50) constant.
In Fig.~\ref{fig:micro_numbuyers_vs_e2e}, we observe that \vader's overhead increases from
a mere 7\% (0.71s) at 100 \MBs\ to 31\% (4.24s) at 500 \MBs.
This is explained by the extra load imposed on the \MFs\
by the signing and verification, steps of \vader.
In contrast, \fullon's e2e time increases rapidly from 3.35s to
30.84s, due to the extra load on the blockchain layer (replicated execute and consensus)
imposed by the increasing no. of commits, from increasing \MB\ count.

\noindent \textbf{\textit{5.2.3} Component Analysis:}
Fig.~\ref{fig:micro_component_vs_buyerload} shows the component contribution
(defined in Metrics, Sec.~\ref{sec:evaluation}) towards e2e
time of a file for \vader\ and \fullon\ protocols with
increasing \#\MBs\ per \MF. 
We observe that for \vader\ there is a 10x increase in network
transfer time (from 0.34s to 3.35s), while protocol time increases 
5x from 0.125s to 0.61s with increasing \MBs. The large network
transfer time increase in \vader\ is attributed to the fact that an
increasing no. of \MBs\ simultaneously fetch content (flash
crowd) thereby putting more stress on \MFs\ and the underlying network. ( We note that \vanilla\
network transfer time behaves in a similar fashion). On the other hand, the protocol time for \fullon\ increases rapidly
(from 2.84s to 29.7s) with increasing \#\MBs\ while the network
transfer time remains the same at around 0.34s. This is due to the 
fact that in the case of \fullon\ \MBs, the blockchain commit wait times 
provide a more spaced out execution and \MF\ network time remains constant.

\begin{figure*}[th!]
    \begin{minipage}[t]{0.285\textwidth}
        \hspace*{-0.2cm}
        \centering
        \includegraphics[scale=0.8,height=1.1in]{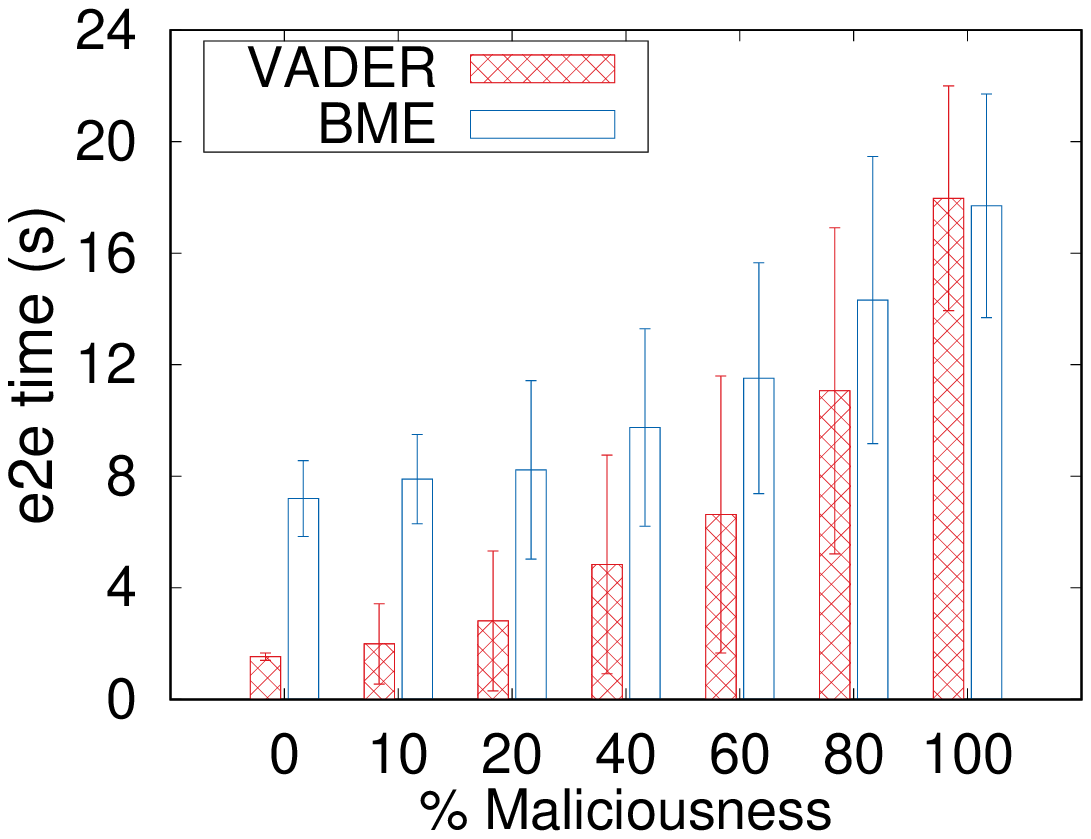}
        \caption{System wide avg e2e time per file vs \% maliciousness (20MB, 50 times)}
        \label{fig:micro_maliciousness_vs_e2e}
    \end{minipage}
    ~
    \begin{minipage}[t]{0.695\textwidth}
        \centering
        \begin{subfigure}[t]{0.298\textwidth}
            \hspace*{-0.8cm}
            \includegraphics[height=1.1in]{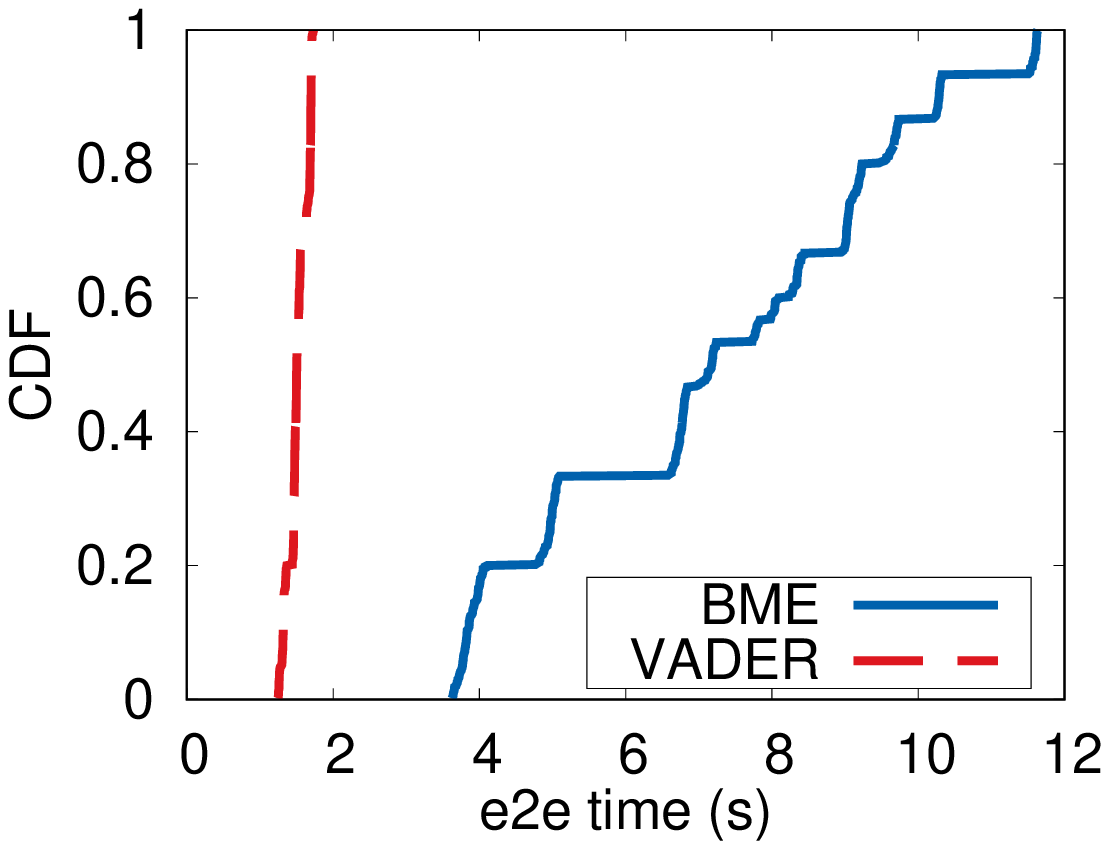}
            \caption{0\% Malicious}
            \label{fig:micro_malicious_cdf_e2e_0_mal}
        \end{subfigure}
    ~
        \begin{subfigure}[t]{0.298\textwidth}
            \hspace*{-0.5cm}
            \includegraphics[height=1.1in]{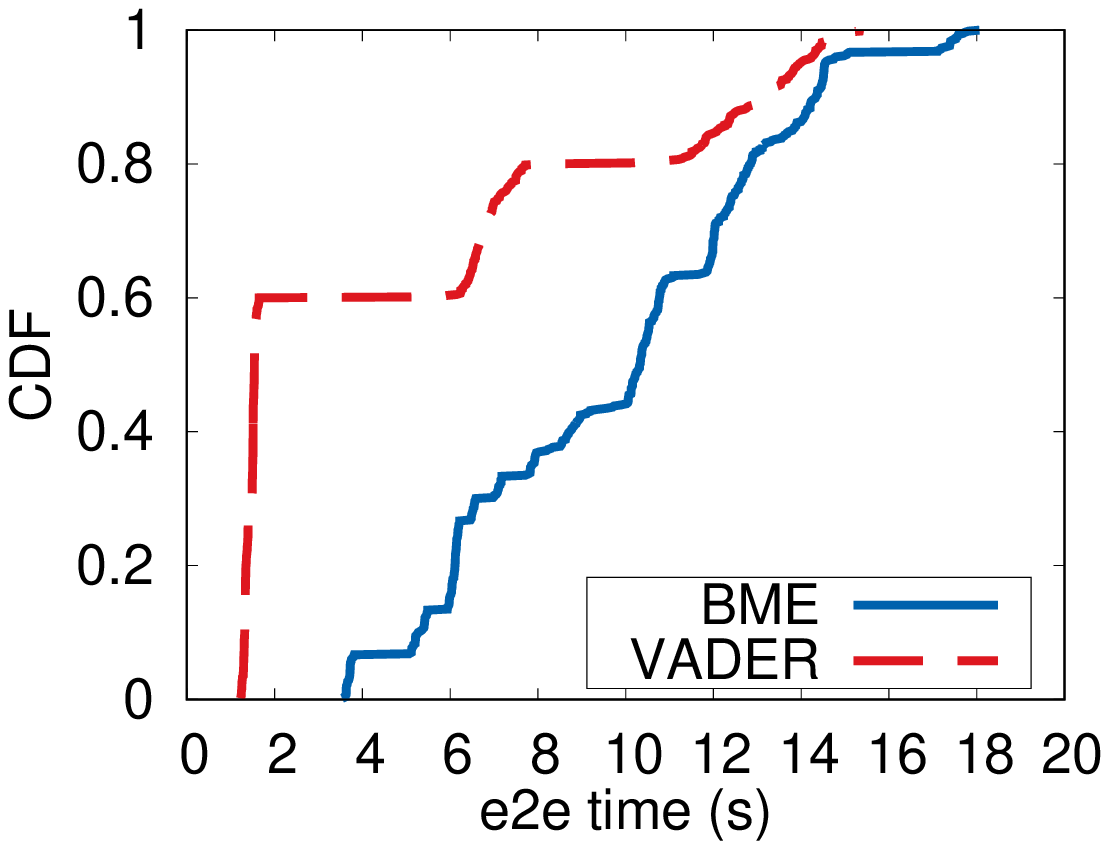}
            \captionsetup{margin={0.3cm,0cm}}
            \caption{40\% Malicious}
            \label{fig:micro_malicious_cdf_e2e_40_mal}
        \end{subfigure}
    ~
        \begin{subfigure}[t]{0.298\textwidth}
            \hspace*{-0.2cm}
            \includegraphics[height=1.1in]{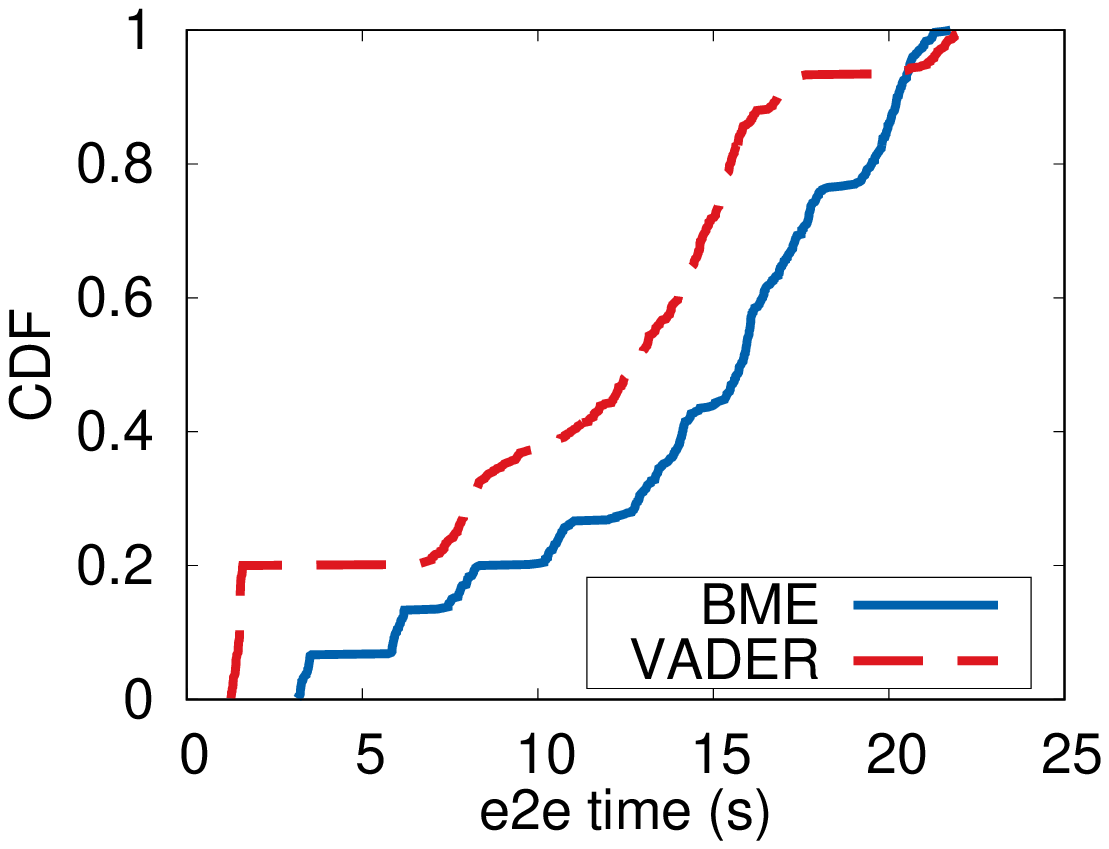}
            \captionsetup{margin={0.7cm,0cm}}
            \caption{80\% Malicious}
            \label{fig:micro_malicious_cdf_e2e_80_mal}
        \end{subfigure}
    \captionsetup{justification=centering}
    \vspace*{-0.3cm}
    \caption{CDF(s) of e2e time per \MB\ under increasing maliciousness. (20MB, 50 times)}
    \label{fig:micro_malicious_cdf_e2e_mal}
    \end{minipage}
\end{figure*}

\noindent \textbf{\textit{5.2.4} Effect of Maliciousness:}
We study the effect of maliciousness by increasing maliciousness from
0\% to 100\% and plot average e2e time for \vader\ and
\fullon\ in Fig.~\ref{fig:micro_maliciousness_vs_e2e}. ~\footnote{
\vanilla\ does not have a dispute resolution mechanism, hence, we don't run \vanilla\ in experiments benchmarking maliciousness.}
We observe that \vader\ starts with lower e2e time (1.53s) compared to \fullon\ (7.2s) for 0\% maliciousness,
while \vader\ e2e time increases with increasing maliciousness and
coming closer to \fullon. This is due to the fact that after the first
dispute, \vader\ closes channel and continues application progress directly on blockchain
similar to \fullon. 

In the worst case of 100\% maliciousness, \vader\ begins to have higher completion time
compared to \fullon\ due to the additional channel open-close overhead.
In Fig.~\ref{fig:micro_malicious_cdf_e2e_mal}, it is interesting to note that
in case of \vader, even in the presence of maliciousness, the non-malicious
parties remain unaffected, and only \MBs\ interacting with a malicious \MF\
incur higher e2e time. For example, e2e time for 60\% of \MBs\ is
less than 1.65s with 40\% maliciousness and jumps to 5.76s right after;
and it is less than 1.59s for 20\% of \MBs\ in 80\% maliciousness,
after which it jumps to 6.34s.

\section{Related Work}
\label{sec:related}
In this section, we review and contrast our work with existing work on building fair, auditable and decentralized systems.

\noindent \textbf{$\bullet$ Blockchain Platforms:}
Beginning with Bitcoin~\cite{nakamoto_bitcoin}, the recent past has seen the emergence of a number of blockchain
platforms~\cite{ethereum,corda,Tendermint,hyperledger_fabric,quorum,zero_cash},
that provide a tamper proof immutable ledger of transactions as well as smart contract capabilities secured
by an underlying consensus protocol.
The various blockchain platforms vary in their choice of consensus protocol,
transaction privacy models, as well as membership in the blockchain network.
Permissioned/consortium model restricts blockchain access to only a few users/organizations, 
while permissionless model allows any node to join and participate in the network.
Typical permissioned blockchain platforms also
support \textit{access control}  capabilities at the smart
contract layer to provide confidentiality guarantees to higher layer
blockchain applications.
\textit{\vader\ relies on granular access control, as well as tamper-proof smart contract execution and auditability of blockchain platforms, to guarantee multi-party fair exchange, even in the presence of passive participants.}

\noindent \textbf{$\bullet$ Blockchain Scalability:}
The underlying consensus protocol overhead limits the scalability of blockchains. Recent solutions to addressing scalability broadly fall in two buckets.
\newline
\noindent \textbf{Layer 1 Solutions:} These solutions involve making
{\it onchain} modifications to the underlying blockchain design such as changing the block size~\cite{bitcoin_cash} or block
generation time~\cite{litecoin}.
The authors in ~\cite{bitcoin_ng} adopt a leader election based
consensus algorithm for achieving faster consensus and block
generation fairness in bitcoin.
~\cite{rscoin, omniledger, elastico, chainspace}
implement sharding to scale transaction throughput wherein the main blockchain is divided
into multiple independent shards within which transactions can be
validated in parallel, and finally merged into the main
chain. Algorand~\cite{algorand} uses a committee based consensus
protocol, while RapidChain~\cite{rapidchain} uses a mix of sharding
and committee based consensus to improve the scaling further.{\it
We note that even if the above works are able to improve
transaction throughput to a few thousand transactions per second,
blockchain consensus will still impose a ceiling on transaction throughput}

\noindent \textbf{Layer 2 Solutions:}
An alternate design to scale throughput is to decouple transaction
processing speed from the underlying blockchain protocol by
reducing the number of interactions with the blockchain. Payment
channels~\cite{poon_lightning,raiden_network} and generalized state
channels~\cite{coleman_counterfactual,perun_network,funfair,spankchain,stefanie_roos_sprites,statechannelnetworks} enable parties to directly transact with each other
off-chain and finally batch multiple transactions into a single
blockchain transaction.
This allows applications to scale independent of the blockchain
protocol while offering security guarantees at par with native blockchain.
While state channels help in scaling, they require \textbf{1)} all parties to be known to each other apriori and \textbf{2)} all parties
(or their delegatee~\cite{pisa}) to continuously remain online. {\it On the other hand, \vader\ works even when parties are not known
  to each other apriori and one of the parties is passive and offline (content owner)}.

\noindent \textbf{$\bullet$ Fair Exchange:}
Fair exchange is a well studied problem, especially fairness for
electronic commerce \cite{asokan_fairness_1998}.
Two party fair exchange without a trusted third party (TTP) is known to be
impossible~\cite{Rabin1981HowTransfer,Goldreich:1987:PAM:28395.28420,YaoExchangeSecrets,FairExchangeInfeasibility}
without relaxing security requirements.
Recent works have looked at
designing blockchain based protocols for fair
exchange~\cite{smc_bitcoin,Kumaresan:2016:ASC:2976749.2978424,Kumaresan:2016:ISC:2976749.2978421,Kiayias:2016:FRM:3081738.3081763}. 
Bentov et.al~\cite{bentov_how_2014} describe a bitcoin based
\textit{claim-or-refund} framework for fair exchange in which either
parties receive the intended goods or are compensated
monetarily. Building on this, the authors in
FairSwap~\cite{FairSwapDziembowskiEF18} describe a framework that
enable a sender and receiver to exchange digital goods such as files
in a fair manner through the use of a `judge' smart contract.
Choudhuri et.al\cite{Choudhuri2017FairnessBoards} design protocols
that ensure fairness in secure multi-party computation using
blockchain-like primitives.
{\it The above protocols guarantee fairness only between the active parties and do not cover passive participants, whereas \vader\ guarantees fairness even for passive parties.}

\noindent \textbf{$\bullet$ Decentralized Marketplaces:}
Recent works on blockchain based decentralized marketplaces allow buyers and sellers to trade directly with each other without a
centralized platform~\cite{Subramanian18, KlemsETHBT17, KabiF18, LBRY, openbazaar}. In these systems, dispute resolution is handled offline in an ad-hoc manner, providing no guarantees on fairness. {\it \vader\ on the other hand guarantees fair exchange and is better aligned with existing centralized marketplaces.}
\newline
\noindent \textbf{$\bullet$ Video Streaming:}
Most commercial video streaming services~\cite{youtube,vimeo}
follow RFC~\cite{rfc_http_streaming} that describes the streaming of
video over HTTP(S).
There has been significant research in the design, measurement and
characterization and optimization of scalable video streaming
systems like~\cite{ConvivaUserEngagement,UnreelingNetflix,VivisectingYoutube,HuluPrefetching,YahooVideoMeasurement,YoutubeCDN,ABRVBR, VideoBitRateAdaptation} characterize the end to end
performance of a commercial video streaming service and identify a
number of bottlenecks across different layers of the
stack. Similarly, \cite{YoutubeCDN} studies policies used for server selection in
the Youtube CDN network. while the authors in~\cite{CDNAugmentation}
study the effects of CDN augmentation
techniques such as P2P-CDN and telco-CDN
federation on video workloads. Similarly, on
the client side, a number of bitrate
adaptation algorithms have been developed
with the goals of minimizing buffering, start
up latency, improving video smoothness etc.
{\it We note that above prior work's focus on improving end user
  experience ard orthogonal to \vader's focus on guaranteeing fairness.}

\noindent \textbf{$\bullet$ Auditing Mechanisms:}
A number of works have looked at auditing running systems.
PeerReview~\cite{peerreview} leverages tamper-evident logging to
detect when a node deviates from the expected
behaviour. AVMs~\cite{auditable_vms} on the other hand, use logging
to record all incoming/outgoing messages from a VM and ensure correct
execution of a remote system.
~\cite{adtributor} helps ad system operators debug revenue problems through multi-dimensional analysis
of various metrics. However, the approach requires access to logs and other internal system metrics. Such works are not applicable in our setting as we do not trust facilitators to act honestly.
Recent works have also focussed on auditing the working of web systems~\cite{UberHood,fiddling_with_pricing,task_rabbit_fairness,taxi_competition, amazon_price_analysis, ecomm_price_discrimination} through black box measurements to detect violations in application defined fairness. {\it  However, these works are limited to detecting unfair practices and do not provide any mechanisms for guaranteeing fairness of the participants.}

\noindent \textbf{$\bullet$ Complementary Technologies:} We highlight complementary technologies that \vader\ can leverage to further enhance the security of the platform. Mangipudi et.al~\cite{kate_watermarking} use a combination of content watermarking and on-chain penalty mechanisms to prevent content piracy which can be easily embedded into \vader\ smart contracts. Emerging technologies such as Intel SGX~\cite{intel_sgx} that provide computational integrity and verifiability through hardware mechanisms could be leveraged for protecting the confidentiality of videos stored on the video server.

\section{Conclusion}
\label{sec:conclusion}
We introduce the problem of {\it Multi-party fair exchange} for digital assets, which requires safeguarding the rights of active and passive participants. We propose a protocol to ensure {\it Multi-party fair exchange} for digital assets, by leveraging blockchain and intelligent incentive alignment. We build a prototype of the protocol on Hyperledger Fabric and extensively evaluate performance of our approach across a realistic test bed and show results that demonstrate the feasibility of our system.


\bibliographystyle{ACM-Reference-Format}
\bibliography{VADER}


\end{document}
\endinput